\documentclass[nofootinbib,superscriptaddress,eqsecnum,tightenlines,11pt,notitlepage,longbibliography]{revtex4-1}
\usepackage[linktocpage]{hyperref}
\usepackage{graphicx}
\usepackage{amsmath,amssymb,amsfonts,amsthm,stmaryrd,mathtools,mathtools}
\usepackage{yfonts}
\usepackage{multirow,bigdelim}
\usepackage{dsfont}
\usepackage{color}
\usepackage{graphicx,color,epstopdf,epsfig,psfrag}
\usepackage[dvipsnames]{xcolor}
\usepackage{ mathrsfs }

\def\be{\begin{equation}}
\def\ee{\end{equation}}
\def\ba{\begin{eqnarray}}
\def\ea{\end{eqnarray}}

\newcommand\nn{\nonumber}
\newcommand{\q}{\quad}

\begin{document}

\title{From 3D topological quantum field theories to 4D models with defects}

\author{Clement Delcamp}
\affiliation{Perimeter Institute for Theoretical Physics,\\ 31 Caroline Street North, Waterloo, Ontario, Canada N2L 2Y5}
\affiliation{Department of Physics $\&$ Astronomy and Guelph-Waterloo Physics Institute \\  University of Waterloo, Waterloo, Ontario N2L 3G1, Canada}
\author{Bianca Dittrich}
\affiliation{Perimeter Institute for Theoretical Physics,\\ 31 Caroline Street North, Waterloo, Ontario, Canada N2L 2Y5}

\begin{abstract}
$(2+1)$ dimensional 
topological quantum field theories with defect excitations are by now quite well understood,  while many questions are still open for $(3+1)$ dimensional TQFTs.  Here we propose a strategy to lift states and operators of a $(2+1)$ dimensional TQFT to states and operators of a $(3+1)$ dimensional theory with defects. The main technical tool are Heegaard splittings, which allow to encode the topology of a three--dimensional manifold with line defects into a two--dimensional Heegaard surface. We apply this idea to the example of BF theory which describes locally flat connections. This shows in particular how the curvature excitation generating surface operators of the $(3+1)$ dimensional theory can be obtained from closed ribbon operators of the $(2+1)$ dimensional BF theory.  We hope that this technique allows the construction and study of more general models based on unitary fusion categories. 
\end{abstract}

\maketitle

\section{Introduction}

In recent years considerable effort has been focussed on understanding topological quantum field theories (TQFTs) with defect excitations. Much progress has been made in understanding topological quantum field theories and their associated defects in $(2+1)$ dimensions, e.g. \cite{Kitaev1,LevinWen,Kong2,Lan,Lan2,Hu,BalKir,Kir}. In particular (Levin-Wen) string nets \cite{LevinWen} provide a huge class of models, each of which is specified by a unitary fusion category. The structure of the excitations of these models is well understood \cite{Kong2,Lan,Hu}. The models are usually given via a Hilbert space description. On this Hilbert space one can define (so-called ribbon) operators, that generate the excitations \cite{Lan,Hu,DG16}.  For a particular class of fusion categories (namely those arising from representation categories of finite groups) the string net models are closely related to Kitaev models \cite{Kitaev1}, more precisely a Fourier (or variable) transformation of the Kitaev models leads to the so--called extended string nets \cite{Bur1, BurKong, BalKir2}. The Kitaev models can be understood to describe the space of flat connections on a 2D surface $\Sigma$ with defect excitations, and it is in this context that ribbon operators have been constructed \cite{Kitaev1}. This set--up is closely related \cite{DDR1,DDR2} to BF theory \cite{horowitz,baezBF} (again with defects), which is a topological field theory describing locally flat connections.  

In contrast, TQFTs in $(3+1)$ dimensions with defect excitations are less developed and understood, see \cite{CM1,CM2,KongWen,Wang,Tiwari} for some recent progress. In the context of $(3+1)$ dimensional BF theory, with a Lie group as structure group, a (lattice and continuum) Hilbert space description together with operators generating and measuring  curvature excitations is detailed in \cite{DG14a,DG14,BDG15}.  However one would also like to have a construction of a Hilbert space and excitation generating operators for the more general case of `q--deformed' BF theory, which  corresponds to string nets for a fusion category associated to a quantum group (at root of unity).   Such a construction has been sketched in \cite{WalkerWang}, but in comparison to the $(2+1)$D models is far less developed and understood. In particular a construction of the operator algebra that would replace the ribbon operators of $(2+1)$D is needed.

The $(3+1)$ dimensional case is particularly important for quantum gravity. Spin foam models \cite{carlobook,baez,perez-review} are one approach to quantum gravity making heavily use of topological field theories, in particular BF theory. A main open problem for spin foam models is the exploration of the large scale limit \cite{Dittrich:2011,BD12, R1,Dittrich:2013,Bahr14,BD14, Miz,BS1,BS2,CDtoappear}. Generically the large scale limit, constructed via coarse graining, is given by some (possibly trivial) topological field theory \cite{Dittrich:2013,CDtoappear}, whereas interacting theories, such as gravity, are expected to arise at phase transitions between these topological field theories \cite{BD12,BD14,BS2}. To understand the dynamics of spin foam models it is therefore important to understand better 4D topological field theories, which could arise via coarse graining from spin foam models,  and their possible (defect) excitations. This will then help to study possible phase transitions. 

As proposed in \cite{DittStein}  4D TQFTs which arise via coarse graining from spin foams and therefore carry a geometric interpretation, can give rise to quantum geometry realizations. That is the TQFT and its defect excitation would lead to a Hilbert space of states, on which an operator algebra can be defined, that admits a geometric interpretation. Thus states of this Hilbert space describe quantum geometries. Importantly this construction leads to diffeomorphism invariant continuum quantum field theories.  In retrospect this has been first realized by the Ashtekar--Lewandowski--Isham construction \cite{ali1,ali2,ali3}, which is based on a trivial topological field theory.  In contrast the $(2+1)$D and $(3+1)$D quantum geometry realization constructed in \cite{DG14a,DG14,BDG15} is based on BF theory. Recently another $(2+1)$D quantum geometry realization was defined \cite{DG16} based on the Turaev-Viro TQFT \cite{TV}, which can be interpreted as q--deformed BF theory.  This realization has many advantages as compared to both  the Ashtekar--Lewandowski--Isham as well as the BF based construction and furthermore provides the dynamics of quantum gravity incorporating a cosmological constant. In particular the quantum field theory is based (via an inductive limit) on a family  of finite dimensional  Hilbert spaces, which makes many subtle regularization procedures  unnecessary.  Thus a $(3+1)$D quantum geometry realization based on q--deformed BF theory would be highly desirable. 

\vspace{5pt} 

We propose in this work a strategy to make many of the techniques developed for the $(2+1)$ dimensional case available for the $(3+1)$ dimensional case.  The main idea is to use a Heegaard splitting \cite{heegardNotes} of the 3D manifold representing a spatial slice of the $(3+1)$ dimensional space--time manifold. Such a Heegaard splitting can be obtained from a triangulation (or lattice) embedded into the 3D manifold. The Heegaard splitting leads to a so--called Heegaard surface. We can then define our operations on this 2D Heegaard surface instead of dealing with the 3D manifold itself.  This allows us to use operators defined for the $(2+1)$ dimensional case in order to describe the $(3+1)$ dimensional one. 

Heegaard splittings and more generally handle decompositions \cite{kirby} have been used to construct topological field theories in 3 and 4 dimensions e.g. \cite{Roberts,BB}. (See also \cite{Ryan} for the use of Heegaard splitting and the associated reduction in dimension in quantum gravity models involving a sum over topologies.)

 Here, however, we deal with topological field theories with defects (i.e. some form of extended topological field theories), in a `canonical' set--up, that is in a Hilbert space description. In this set-up we are in particular looking for operators generating and measuring the defect excitations.  We will describe how such operators can be induced for the $(3+1)$ dimensional case from the $(2+1)$ dimensional theory.


To make the main idea clear we will focus in this work on BF theory associated to a finite group and leave the q--deformation and more general string net models for future work. That will allow us to speak of the space of flat connections and to make intuitive use of its properties. 

~\\

The paper is structured as follows: In section \ref{2Dstuff} we will review the construction of a Hilbert space describing locally flat connections of a 2D surface and (ribbon) operators defined on such a Hilbert space.  The ribbon operators can generate curvature for non--contractible cycles of the 2D surface and in this sense generate excitations. In section \ref{3Dstuff} we will then describe how to encode the space of flat connections on a 3D manifold with defects into the space of flat connections on a 2D surface. This 2D surface is a Heegaard surface defined by the structure of the defects. In our case the defects are confined to a triangulation and thus the Heegaard surface is specified by this triangulation as well.  A 2D flat connection needs to satisfy additional flatness constraints in order to describe a (defected) 3D flat connection. Accordingly we will discuss which class of operators defined in the 2D setting, preserve these constraints. This will lead to (excitation generating) operators for the 3D case.  In section \ref{4simplex} we will discuss in much detail an example triangulation and the corresponding (ribbon) operators. In sections \ref{2Dstuff} to \ref{4simplex} we will focus on Hilbert spaces of gauge invariant functions and gauge invariant operators. In section \ref{torsion} we will shortly discuss relaxing gauge invariance and allow another class of excitations, known as torsion (in the gravitational context) or electric excitations.  We close with an outlook in section \ref{outlook}.

\section{Space of flat connections and ribbon operators on 2D surfaces}\label{2Dstuff}

In this section we are going to construct  Hilbert spaces of functions of locally flat connections on 2D surfaces. We will furthermore  describe the operators on this Hilbert space and their action. Later-on we will use these Hilbert spaces and operators to describe the space of locally flat connections with curvature defects on 3D manifolds.

\subsection{Configuration space / Hilbert space}

Consider a closed 2D manifold $\Sigma$. We are interested in the space of flat ${\cal G}$--connections on this surface. For simplicity we will assume here that ${\cal G}$ is a finite group (we will comment in section \ref{torsion} about generalizations to Lie groups. The space of flat connections on $\Sigma$ can be captured by choosing a  directed connected graph $\Gamma$ embedded in $\Sigma$, so that the cycles of $\Gamma$ capture (at least) the first fundamental group $\pi_1(\Sigma)$. 
One can then restrict the attention to the first fundmamental group $\pi_1(\Gamma)$ of the graph $\Gamma$ equipped with additional constraints, that demand that every contractible cycle in $\Sigma$ carries trivial holonomy. This can be expressed as 
\ba\label{2.1}
\{ h \in  \text{Hom} (\pi_1(\Gamma),{\cal G}) \,  | \,  h(c)=\mathbb{I}  \q \forall c \,\, \text{contractible in $\Sigma$}\} /  {\cal G}
\ea
where we quotient out by the diagonal adjoint action of ${\cal G}$. (That is, one has chosen some root node among the nodes of $\Gamma$ which serves as source and target nodes for the cycles of $\pi_1(\Gamma)$. The group ${\cal G}$ acts by gauge transformations on this root and induces the diagonal adjoint action.)

We want to construct a Hilbert space of functions on the space of flat connections on $\Sigma$. Such functions can be expressed as follows:  We first consider the space ${\cal F}$ of functions $\psi: {\cal G}^L \rightarrow \mathbb{C}$  where $L$ denotes the number of links $l$ in the graph $\Gamma$.  A graph connection is given by the assignment of group elements $g_l$ to the (directed) links $l$ of the graph $\Gamma$.  

To ensure that a function $\psi\in {\cal F}$ defines a consistent function on the space $\text{Hom} (\pi_1(\Gamma),{\cal G})$ we have to impose gauge invariance. (If we include gauge invariance at the root node we will also impose invariance under the diagonal adjoint action in \eqref{2.1}.)
The gauge transformations act at the nodes $n$ of the graph $\Gamma$, whose number is denoted by $N$. A gauge transformation is parametrized by $\{u_n\}_n \in {\cal G}^N$  and acts on a graph connection $\{g_l\}_l$ as
\ba
\{u_n\}_n \triangleright  \{g_l\}_l \,=\, \{u^{-1}_{t(l)} \,g_l \, u_{s(l)} \}_l \,   \q ,
\ea
where $t(l)$ denotes the target node of the link $l$ and $s(l)$ its source node.
The space of gauge invariant functions, denoted by  ${\cal F}_{\bf G}$ is the space invariant under this action for all $\{u_n\}_n \in {\cal G}^N$. We can encode this  into (stabilizer) constraints
\ba
({\bf G}_{\{u\}} \psi)\{g\}  &   \,:= \, & \psi\{   u^{-1}_{t(l)} \,g_l \, u_{s(l)}    \} \,\stackrel{!}{=}\, \psi(g)  \q .
\ea

  Each function $\psi \in {\cal F}_{\bf G}$ defines a function $\Psi$ on $\text{Hom} (\pi_1(\Gamma),{\cal G})$ by setting
\ba
\Psi( \{ h(c)\}_{c\in \pi_1(\Gamma)}) &=& \psi( \{ g_l\}_{l \in \Gamma})
\ea
where $\{g_l\}_l$ is a graph connection so that $h(c)\,=\, g_{l_n} \cdots g_{l_1}$ for every cycle $c=l_n \cdot \ldots \cdot l_1$ of the graph. The choice of the particular gauge connection $\{g_l\}_l$ satisfying this condition does not matter due to the gauge invariance condition. 

We need furthermore to impose the (flatness) constraints reducing  $\text{Hom} (\pi_1(\Gamma),{\cal G})$ to the space of flat connections on $\Sigma$. 
Thus for every independent cycle $c= l_{n} \circ \cdots \circ  l_{1}$ of $\Gamma$ that is contractible in $\Sigma$ we impose the condition 
\ba
({\bf F}_c \psi)\{g\} \,:=\, \delta(\mathbb{I}, g_{l_n} \cdots g_{l_1}) \, \psi\{g\} \, =\, \psi\{g\}   \q .
\ea
  (Note that the flatness constraints are invariant under the gauge action and can therefore be imposed on ${\cal F}_{\bf G}$.)  This defines ${\cal F}_{\bf G,\bf F}$, the space of functions in ${\cal F}_{\bf G}$ satisfying all flatness constraints imposed by $\Sigma$--contractible cycles.  Such functions define states on the configuration space given by the first fundamental group $\pi_1(\Sigma)$.

On all spaces ${\cal F}_{\bullet}$ we can consider the inner product given by 
\ba\label{innerpr}
\langle  \psi_1 \,|\, \psi_2   \rangle  &=& \frac{1}{|{\cal G}|^L} \sum_{g_l} \overline{\psi_1} \{g_l\} \,\, \psi_2 \{g_l\}  \q 
\ea
and thus obtain the corresponding Hilbert spaces ${\cal H}_{\bullet}$.

\subsection{ Ribbon operators}\label{ribbonop}

Next we will discuss operators on  the Hilbert space ${\cal H}_{{\bf G},{\bf F}}$ of gauge invariant functions on the space of flat connections on $\Sigma$.  We will express these operators as  operators on ${\cal F}$,  the space of functions on graph connections. But the operators have to commute with the constraints ${\bf F}_c$ and ${\bf G}_{\{u\}}$, which specify the space ${\cal H}_{{\bf G},{\bf F}}$.

One class of operators  are given by Wilson loop or (closed) holonomy operators $W^f_\gamma$ which act by multiplication on states $\psi\{g\}$.  To begin with let us assume that $\gamma$ coincides with some closed path along the links of $\Gamma$. The (gauge invariant) holonomy operator $W^f_\gamma$ is parametrized by a class function $f$ on the group ${\cal G}$ and its action is given by
\ba
(W^f_\gamma \psi) \{g\} \,=\, f(h_\gamma) \, \psi \{ g\}
\ea
where $h_\gamma = g_{l_n} \cdots g_{l_1}$ if $\gamma=l_n \circ \cdots \circ l_1$.

 Note however that as we are working with the space of flat connection, we can deform the path $\gamma$ without changing the operator, that is $W^f_\gamma$ only depends on the isotopy class of $\gamma$ in $\Sigma$.   Thus we can allow also paths $\gamma'$ not coinciding  with some subset of links in $\Gamma$, as we can replace such $\gamma'$ with some isotopy equivalent $\gamma$ which runs on the graph $\Gamma$.

The second class of operators are known as (exponentiated) flux operators in loop quantum gravity, and act as translations on the arguments of the wave functions, that is as exponentiated derivative operators (see \cite{BDG15} for a precise definition for the Lie group case). 

Such  translation operators  violate  in general the flatness constraints: A state satisfying the flatness constraint for a cycle $c$  is of the form
\ba
\psi\{g\} \,=\, \delta(\mathbb{I}, h_c)\, \psi\{g\} \q .
\ea
This form would be destroyed if one of the group elements $g_{l_i}$ appearing in $h_c=g_{l_n} \cdots g_{l_1}$ is translated.  The idea is to define  operators that combine a number of translations such that all holonomies associated to contractible cycles stay invariant.  

This is the main idea behind the ribbon operators of Kitaev \cite{Kitaev1} and leads also to the definition of so--called integrated flux operators in \cite{DG14a,DG14,BDG15}. In the context of $(2+1)$ gravity such operators (assuming also gauge invariance can be preserved) are Dirac observables of the theory, i.e  operators that commute with the constraints \cite{FreiZap}.  

We are now going to define the translation operators $T^H_{\alpha}$  based on a closed directed and non--intersecting path $\alpha$. Via the process of gluing (open) ribbons one can also deal with intersecting paths, but for this one needs to specify which parts of the path are over-- or under--crossing. See e.g. \cite{Hu,DG16,DDR1}

 To begin with we assume that the path $\alpha$ is such that it intersects links of $\Gamma$ transversally and does avoid the nodes of the graph. (In the end the operator only depends on the isotopy class of $\alpha$.) We also assume that all crossing links have the same orientation with respect to $\alpha$: to be specific, if the direction of $\alpha$ is upwards the links should cross from left to right. Note that  by a variable transformation $g_{l^{-1}}=g_l^{-1}$ for the wave functions we can always adjust the orientation of the links accordingly.   We furthermore need to specify one link $l_1$ crossed by $\alpha$ as initial link. Again it will turn out that this choice does not matter for the final (gauge invariant) operator.

To the path $\alpha$ we associate a `shadow' path $\alpha'$ which has to be isotopy equivalent to $\alpha$ and to run along $\Gamma$, i.e. $\alpha'$ is composed from links of $\Gamma$. More specifically $\alpha'$ has to connect the target nodes of the links $l \in \Gamma$  which are crossed by $\alpha$ and is  not allowed to cross $\alpha$ itself.  Note that even if these condition do not seem to be satisfiable for a given path $\alpha$ and a given graph $\Gamma$ one can refine $\Gamma$ to obtain an appropriate $\alpha'$. We will discuss an example of this procedure in appendix \ref{refiningG}. 
With these definitions we can picture $\alpha$ and $\alpha'$ as the left and right boundary of a ribbon respectively. Thus the right boundary of this ribbon needs to be aligned to links of the graph $\Gamma$.

To define the action of the translation operator $T^H_\alpha$ associated to this ribbon we denote  the links crossed by the closed path $\alpha$ by $l_i,\, i=1 \ldots, l_{N(\alpha)}$ and the associated holonomies by $g_i$. 
Furthermore $h'_{i} := h_{t(l_1)t(l_i)}$ is the holonomy from the target node of the link $l_i$ along and in the direction of $\alpha'$ to the target node of $l_1$. For $i=1$ we define $h'_{1}=h_{\alpha'}$ to be the holonomy of $\alpha'$ starting and ending at the target node of $l_1$.

\begin{figure}[!]
	\centering
	\begin{minipage}[b]{1\textwidth}
		\centering
		\includegraphics[scale = 1]{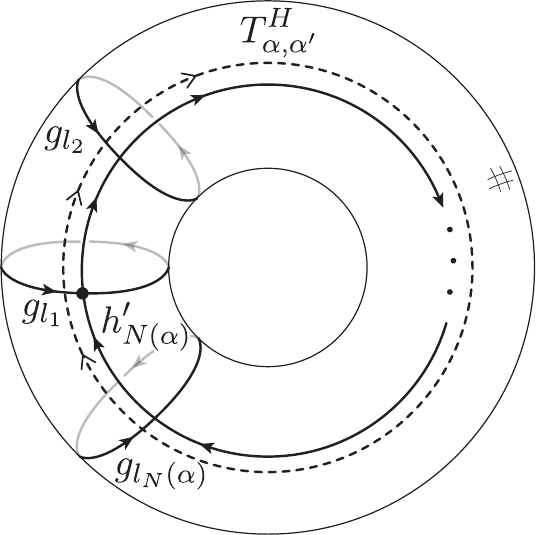}	
	\end{minipage}
	\caption{  A ribbon operator along a non--contractible cycle on a torus. \label{cribbon}}
\end{figure}

We can then define the action of the operator $T^H_{\alpha,\alpha'}$ as
\ba\label{defT1}
(T^H_{\alpha} \psi)\{g\} &=& \psi(   \{   (h'_i)^{-1} H^{-1} h'_i  g_{l_i} \}_{l=l_i}\, , \{ g_{l}\}_{l\neq l_i})  \q .
\ea

This operator leaves almost all flatness constraints for the $\Sigma$--contractible cycles intact. The exceptional cycles result from the possibility that the operator might violate the flatness condition for one face of $\Gamma$, namely the face into which $\alpha$ enters via crossing the `last' link $l_{N(\alpha)}$.  Thus $\alpha$ leaves the same face by crossing $l_1$.  The action of the shifts on two links usually cancels out for the face holonomy, but in this case one might encounter a non--trivial result 
as the holonomy $h_f$ is changed by
\ba
h_f  =  \ldots     g_{l_{1}}^{-1}   h'_{N(\alpha)}  g_{l_{N(\alpha)}}     \q \rightarrow \q
   \ldots    g_{l_{1}}^{-1}   (h'_1)^{-1} H h'_1    H^{-1}  h'_{N(\alpha)} \, g_{l_{N(\alpha)}} \q .
\ea
This is due to the fact that the parallel transport for the shift of $l_1$ involves $h'_1=h_{\alpha'}$, that is the holonomy associated to the cylce $\alpha$, whereas the parallel transport for $l_{N(\alpha)}$ only involves $h'_{N(\alpha)}$, the holonomy from the target node of $l_{N_(\alpha)}$ to the target node of $l_1$.

Flatness at such a face is preserved if $H$ and $h_{\alpha'}$ do commute. One way to ensure this is to combine the translation operator $T^H_{\alpha}$  with a holonomy operator $W^f_{\alpha'}$ with $f(\cdot) = \delta(G, \cdot)$ where $G$ and $H$ have to commute: $GH=HG$. Note that now $f(\cdot)= \delta(G, \cdot)$ is,  for $G$ not in the centre of ${\cal G}$,  not a class function anymore and thus $W^f$ violates in general gauge invariance at the target vertex of $\alpha'$. But later--on we will apply a gauge averaging procedure for the combined operators which will recover gauge invariance.

The combined operators define the so--called closed ribbon operators \cite{Kitaev1,Bombin,DDR1} 
\ba\label{rib1}
{\cal R}_{\alpha} [G,H]  &=&    \delta(GH,HG)  \, W_{\alpha'}^{\delta(G,\cdot)}\, T^H_{\alpha} \q .
\ea
If $G \neq \mathbb{I}$ and $\alpha$ (and therefore $\alpha'$) is a contractible cycle, the ribbon operator (\ref{rib1}) annihilates all states in ${\cal H}_{\bf F}$. If  $G = \mathbb{I}$ and $\alpha$  is a contractible cycle, the ribbon operator will act as identity.

We also need to insure that the ribbon operator preserves gauge invariance. In the construction of the closed ribbon we use the node $t(l_1)$ as defining a reference system to which the translation parameter $H$ is transported. Furthermore we have $h_{\alpha'}=G$ starting and ending at $t(l_1)$. Thus gauge invariance will be preserved at all nodes except $t(l_1)$, where it might be violated.

To regain gauge invariance at this node we can apply an averaging over the gauge group, acting at this node, to the wave function resulting from the action of the ribbon operator. This gives
\ba
&&  \sum_{u \in {\cal G}}   \delta(G,  u h_{\alpha'} u^{-1} ) \,\,    \psi(   \{   (h'_i)^{-1}  u^{-1} H^{-1} u h'_i  g_{l_i} \}_{l=l_i}\, , \{ g_{l}\}_{l\neq l_i})  \nn\\
&=&  \sum_{u \in {\cal G}}   \delta( u^{-1}G u ,   h_{\alpha'} ) \,\,    \psi(   \{   (h'_i)^{-1}  u^{-1} H^{-1} u h'_i  g_{l_i} \}_{l=l_i}\, , \{ g_{l}\}_{l\neq l_i}) \nn\\
&=& \sum_{u \in {\cal G}}  \left( {\cal R}_\alpha [u^{-1} G u, \, u^{-1} H u]  \, \psi \right)( \{g_l\}_l ) \nn\\
&=:&  \left( {\cal R}_\alpha [D, \, C]  \, \psi \right)( \{g_l\}_l ) \q .
\ea
Here we used in the first line that the initial state $\psi$ is gauge invariant, in the second line we rewrote the gauge averaging on arguments of the wave functions as a gauge averaging of the parameters $G,H$ of the ribbon. This defines (in the last line) an operator that only depends on the conjugacy class $C$ (of $H$) and on the conjugacy class $D$ (of $G$) in the stabilizer group $N_H$ of $H$ in ${\cal G}$.  (Remember that $G$ has to commute with $H$, so $G$ is in the stabilizer group $N_H$. Furthermore we can represent $u$ as $u=qn$ with $n\in N_H$ and $q \in {\cal G}/N_H$. The sum over $q \in {\cal G}/N_H$ reduces the information about $H$ to the conjugacy class $C$ and the sum over $n\in N_H$ reduces the information about $G$ to the conjugacy class $D$.)  The group averaging also removes the dependence of the choice of a `first' link $l_1$ among the links crossed by $\alpha$.

We have seen that the ribbon operator preserves flatness for the contractible cycles. It can thus change only the holonomies associated to non--contractible cycles, which are crossed by $\alpha$.  Furthermore one can see that the action of the ribbon on the cycle holonomies  is invariant under isotopic deformations of $\alpha$ and thus the ribbon operator does only depend on the isotopy class of $\alpha$. 

Ribbon operators with $G=\mathbb{I}$  (or $D=\{\mathbb{I}\}$) with $\alpha$ non--contractible will play a special role in the later discussion. In this case there is no restriction on $H$ or on the conjugacy class $C$.

\section{ 3D set-up} \label{3Dstuff}

In this section we are going to explain how to use the Hilbert spaces and operators describing a theory of locally flat connections on a 2D surface in order to obtain a Hilbert space and operators   for a theory of flat connections and curvature defects on a 3D manifold.

\subsection{ Heegaard splittings of 3D manifolds based on a triangulation}

To this end we need to review some basics on Heegaard splittings \cite{heegardNotes} of 3D manifolds.

Let ${\cal M}$ be a compact, closed, connected, orientable 3D manifold. Such manifolds allow for Heegaard splittings. A Heegaard splitting is defined by a triple $(\Sigma, {\cal M}_1, {\cal M}_2)$ where ${\cal M}_i,i=1,2$ are handlebodies  in ${\cal M}$ such that ${\cal M}_1 \cup {\cal M}_2={\cal M}$ and $\Sigma$ is a closed surface embedded in ${\cal M}$, called the Heegaard surface, such that $\partial {\cal M}_1 =\partial {\cal M}_2 = \Sigma$.

Handlebodies are 3D manifolds with boundaries that arise from gluing  a collection of closed 3--balls. This gluing is specified by choosing pairs of disks $D,D'$ on the boundaries of the 3-balls that are identified with each other through a choice of homeomorphisms  $\phi:D\rightarrow D'$.

Given ${\cal M}$ we can obtain a Heegaard splitting through a triangulation of ${\cal M}$:   A regular neighbourhood  of the 1--skeleton of the triangulation (that is the set of vertices and edges of the triangulation) defines ${\cal M}_1$ and the complement of ${\cal M}_1$ in ${\cal M}$ defines ${\cal M}_2$.

${\cal M}_1$ can be understood as a blow--up of the one--skleleton $\Delta_1$ of the triangulation: the blown up vertices define punctured 3--balls, where the punctures arise through the (blown up) edges  adjacent to the vertices. Gluing the 3--balls along the punctures as prescribed by the edges gives the handlebody ${\cal M}_1$.

We want to consider the configuration space of flat connections on ${\cal M} \setminus \Delta_1$ which is equivalent to the space of flat connections on ${\cal M}_2$. This space of flat connections on ${\cal M} \setminus \Delta_1$ allows to have curvature defects concentrated on the edges of the triangulation, that is on $\Delta_1$. 

We aim at mapping the space of flat connections on ${\cal M}_2$ to the space of flat connections on the Heegaard surface $\Sigma(\Delta)$, obtained from using the triangulation $\Delta$ to define the Heegaard splitting. 
Clearly, a flat connection on ${\cal M}_2$ is determining a flat connection on $\Sigma(\Delta) = \partial {\cal M}_2= \partial {\cal M}_1$. 

\begin{figure}[!]
	\centering
	\begin{minipage}[b]{1\textwidth}
		\centering
		\includegraphics[scale = 1]{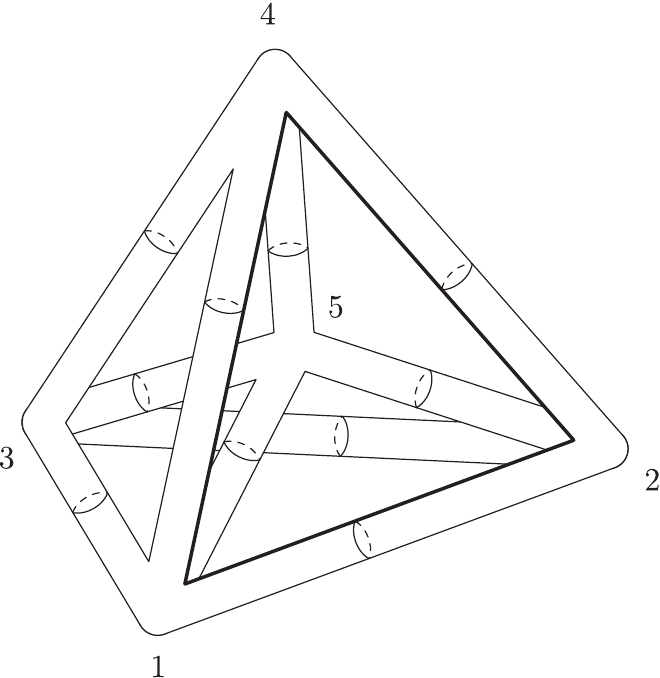}	
	\end{minipage}
	\caption{The Heegaard surface induced by the triangulation of the 3--sphere with the boundary of a 4--simplex.  We label the vertices of the triangulation and the associated punctured spheres of the Heegaard surface by numbers 1, \ldots 5.  The bold line indicates the curve $t \cap \Sigma(\Delta)$ for a triangle $t$ with vertices $v=1,2,4$.  \label{Fsimplex}}
\end{figure}

But not all flat connections on $\Sigma(\Delta)$ can be understood as flat connections on ${\cal M}_2$. 
In fact there are non--contractible cycles in $\Sigma(\Delta)$ that can be contracted to a point in ${\cal M}_2$.  In the case that we arrived at the Heegaard splitting through a triangulation, a (over--) complete basis of such cycles is given by the triangles of the triangulation in the following way: 

  For a given triangle $t$ the curve given by  the intersection $t \cap \Sigma(\Delta) $  defines a cycle on $\Sigma(\Delta)$ that is contractible in ${\cal M}_2$ but in general not contractible in $\Sigma(\Delta)$.   (See figure \ref{Fsimplex} for an example of such a curve.) Thus for every triangle $t$ we have to impose the constraint that the holonomy  associated to the curve $t \cap \Sigma(\Delta) $ is trivial. 
  
  The question is then whether the set of constraints arising from the set of all triangles is sufficient to ensure that a flat connection on $\Sigma(\Delta)$ defines a flat connection on ${\cal M}_2$.  This sufficiency follows from the fact that the set of triangles defines a so--called `system of disks'  \cite{heegardNotes} for ${\cal M}_2$: that means that cutting ${\cal M}_2$ along  all the cut sets $t \cap {\cal M}_2$ we remain with a set of 3--balls. Indeed in our set--up the 3--balls are in one--to--one correspondence with the tetrahedra of the triangulations.  The boundaries of these 3--balls can be seen as punctured spheres with the punctures given by the cut sets $t \cap {\cal M}_2$. For each 3--ball the curves defined by the boundary of these punctures, that is by the sets $t\cap \Sigma(\Delta)$, give a (over)--complete set of curves that are contractible in the bulk of the 3--ball, but not necessarily on the punctured 2--sphere itself. Gluing back the 3--balls to ${\cal M}_2$ we see that indeed taking the curves defined by $t\cap \Sigma(\Delta) $ for all triangles $t$ in the triangulation, gives an (over--) complete set of cycles on $\Sigma(\Delta)$ that are contractible in ${\cal M}_2$.

  

As we want to represent the space of flat connections in ${\cal M}_2$ we have to demand that the holonomies associated to these cycles are flat. 
In this way we obtain a set of flatness conditions $\{{\bf F}_t\}_{t\in \Delta}$, one for each triangle of the triangulation. A locally flat connection on $\Sigma$, satisfying these flatness constraints, defines also a locally flat connection on ${\cal M}_2$ and thus on ${\cal M}\setminus \Delta_1$.

\subsection{ Flat connections on the Heegaard surface}\label{graph}

Thus the space of flat connections  on ${\cal M}\setminus \Delta_1$ can be identified with the space of flat connections on $\Sigma(\Delta)$  with flatness conditions $\{ {\bf F}_t\}_{t\in \Delta}$ imposed.

To describe this space we construct a graph $\Gamma$ in $\Sigma(\Delta)$. This graph will be not minimal, that is links can be removed without loosing information.  It will however be more convenient to work with a non--minimal graph. In the following we will define this graph $\Gamma$:

To begin with we describe the structure of the surface $\Sigma(\Delta)$. A vertex $v$ of the triangulation $\Delta$, with $N_v$ adjacent edges (if there are edges with coinciding source and target vertices one needs to count them twice),   is surrounded by an $N_v$--punctured sphere, which is part of $\Sigma(\Delta)$. We will denote the sphere associated to a vertex $v$ by ${\cal S}_v$. The punctures of ${\cal S}_v$ are due to the edges of the triangulations which are surrounded by tubular pieces ${\cal T}_{v'v}$ of $\Sigma(\Delta)$. The punctures are given by disks that are removed from the sphere and along which the tubular pieces are glued. A tubular piece is also a two--punctured sphere. Note that for making the gluing well defined we need to mark a point on the boundary of each puncture: two punctures are then glued to each other by matching the marked points.  

We can now describe the graph $\Gamma$ on $\Sigma(\Delta)$ by specifying it on each of the pieces ${\cal T}_{vv'}$ and ${\cal S}_v$.

The tubes ${\cal T}_{vv'}$ will carry just one link of the graph $\Gamma$, connecting the marked point of one puncture with the marked point of the other puncture. We demand that the marked points on the punctures are in the cut set $t \cap {\cal T}_{vv'}$ for one of the triangles $t$ adjacent to the edge $e(vv')$ connecting $v$ and $v'$. The link of the graph $\Gamma$ along the tube ${\cal T}_{vv'}$ is then required to be isotopic to the curve induced by this triangle $t$ in ${\cal T}_{vv'}$, that is to the intersection $t \cap {\cal T}_{vv'}$. This requirement avoids a winding of the links around the tubes, as compared to the curves $t \cap \Sigma(\Delta)$.  We call $g_{vv'}$ the holonomy associated to the link of the graph on the tube  ${\cal T}_{vv'}$, going from $v'$ to $v$.

\begin{figure}[!]
	\centering
	\begin{minipage}[b]{1\textwidth}
		\centering
		\includegraphics[scale = 1]{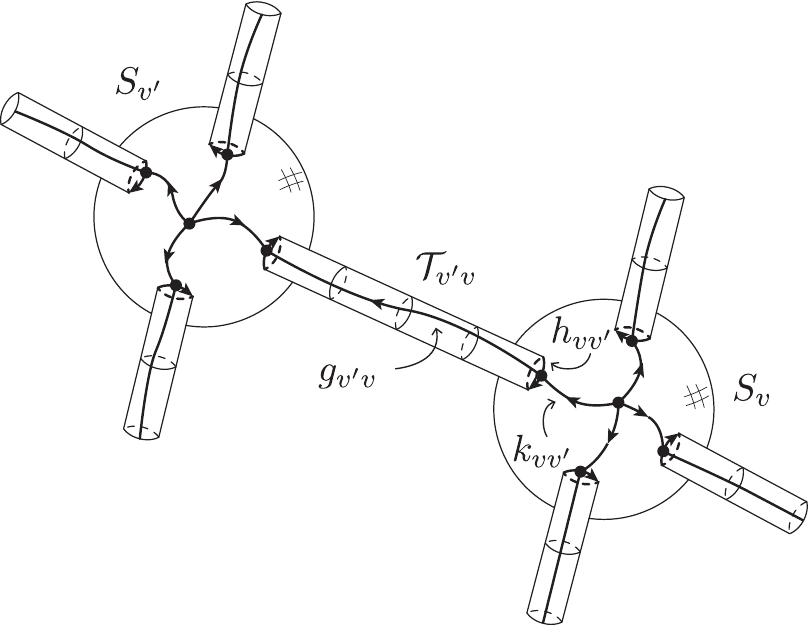}	
	\end{minipage}
	\caption{  This figure shows a part of a Heegaard surface consisting of two punctured spheres connected by a tube. It also shows the graph $\Gamma$ on this part of the surface. On the spheres we have links surrounding the punctures as well as links running from a central node to the punctures. There is also a link along each tube that connects the pieces of graphs on  different spheres. \label{Hspheres}}
\end{figure}

For a sphere ${\cal S}_v$  we choose a position for an $N_v$--valent node $n_v$ of the graph $\Gamma$ and connect this node by non--intersecting links to all the marked points on the boundary of the disks, representing the $N_v$ punctures, see figure \ref{Hspheres}.  We denote by $k_{vv'}$  the holonomy from $n_v$ to the marked point of the puncture to which the tube ${\cal T}_{v'v}$ is glued.
We furthermore surround each puncture with a cycle which starts and ends in an additional node placed on the link connecting $n_v$ to the puncture in question. To avoid the introduction of further holonomies, we take the limit where this additional node coincides with the marked point on the boundary of the puncture.  The  clockwise oriented holonomy around the cycle surrounding the puncture glued to ${\cal T}_{v'v}$ (and on the sphere ${\cal S}_{v})$ will be named $h_{vv'}$.

Connecting all punctured spheres ${\cal S}_v$ with tubes ${\cal T}_{vv'}$ according to the connectivity of the vertices in the triangulation we reconstruct the surface $\Sigma(\Delta)$ and an embedded graph $\Gamma$. 

We can therefore consider the space of connections on $\Gamma$. To give a flat connection on $\Sigma(\Delta)$ satisfying the triangle--flatness constraints, a graph connection needs to satisfy a number of conditions. In the following we list again the holonomy variables and these conditions:
\ba
&&k_{vv'} \q\q \text{holonomy  on ${\cal S}_v$ from $n_v$ to the puncture glued to ${\cal T}_{v'v}$}\nn\\
&&h_{vv'}\q\q\text{holonomy  on ${\cal S}_v$ circling  clockwise the puncture glued to ${\cal T}_{v'v}$} \nn\\
&&g_{v'v}\q\q\text{holonomy on ${\cal T}_{v'v}$ from the puncture glued to ${\cal S}_v$ to the puncture glued to ${\cal S}_{v'}$} .\q\q\nn
\ea
The constraints on the holonomies are given by:
\begin{itemize}
\item For each tube ${\cal T}_{vv'}$ we have one constraint that fixes one of the two cycle holonomies surrounding the cylinder in terms of the other  cycle holonomy and the holonomy going along the tube:
\ba
h_{vv'} \,=\,  g_{vv'}  h_{v'v}^{-1}  g_{v'v}
\ea 
(Note that $g_{vv'}=g_{v'v}^{-1}$ according to our definitions.)
\item For each sphere ${\cal S}_v$ we have one flatness constraint (actually the integrated Bianchi identity), that involves the cycles $h_{vv'}$ around all punctures of the sphere parallel transported to $n_v$.  Assuming that  $D_{vv_1}, D_{v_2}, \cdots  D_{v_{N_v}}$ is a clockwise ordering of the punctures $D_{vv'}$ on $S_v$ if looking from $n_v$ we have the constraint
\ba
(k^{-1}_{vv_{N_v}} h_{vv_{N_v}}k_{vv_{N_v}}) \cdots  (k^{-1}_{vv_2} h_{vv_2}k_{vv_2}) \,\, (k^{-1}_{vv_1} h_{vv_1}k_{vv_1})\,=\, \mathbb{I}  \q .
\ea
\end{itemize}

These two types of constraints ensure that the connection on the Heegaard surface $\Sigma(\Delta)$ is locally flat. In addition we have the triangle flatness constraints, ensuring that the connection on the Heegaard surface defines a locally flat connection on ${\cal M}\setminus\Delta_1$:
\begin{itemize}
\item For each triangle $t$ in the triangulation $\Delta$ we have a constraint ${\bf F}_t = \mathbb{I}$.  Note that the previous two classes of constraints are just arising from having a flat connection on the surface $\Sigma(\Delta)$. In contrast to that the constraints $\{{\bf F}_t\}_{t\in \Delta}$ ensure that a flat connection on $\Sigma(\Delta)$, that satisfies these constraints, determines a flat connection on ${\cal M}\setminus \Delta_1$.  A constraint ${\bf F}_t$ will involve  holonomies $k_{vv'}$ and $g_{vv'}$, but depending on one choices for the graph $\Gamma$ also holonomies $h_{vv'}$ might appear. We will give examples in section \ref{4simplex}. 
 \end{itemize}

We can again consider the space of gauge invariant (wave) functions on the space of flat connections on $\Sigma(\Delta)$, which also need to  satisfy the additional constraints $\{{\bf F}_t\}_{t\in \Delta}$. We equip this space with a Hilbert space structure by choosing the usual inner product (\ref{innerpr}).  The  resulting Hilbert space is denoted by ${\cal H}_\Delta$.

States of the form
\ba
\psi_{\rm vac}= \prod_{e} \delta( h_{s(e)t(e)}, \mathbb{I}) \, \delta( h_{t(e)s(e)}, \mathbb{I}) \,\, \prod_t \delta( {\bf F}_t, \mathbb{I}) \,\, \psi
\ea
satisfy all the flatness constraints and have also vanishing curvature around all edges of the triangulation. (Here $e$ denotes an edge of the triangulation and $s(e), t(e)$ its source and target vertex respectively.) Such states  can therefore be taken as vacuum states. Here the degeneracy of the vacuum might result from non--contractible cycles in ${\cal M}$, that can carry non--trivial holonomy. This is encoded in the function $\psi$, assumed to be a (gauge invariant) functional of these holonomies.

\subsection{Operators on the space of flat connections on ${\cal M}\setminus \Delta_1$}

In section \ref{ribbonop} we discussed operators on the space of flat connections on a surface $\Sigma$. We now aim at finding operators in ${\cal H}_\Delta$, that is operators acting on functions on the space of flat connections on $\Sigma(\Delta)$ which in addtion leave the triangle-flatness constraints invariant. 

One class of operators, for which this holds are Wilson loop operators: these act by multiplication on the function space of holonomies and thus commute with any flatness constraints.

Another class is given by a special kind of (gauge averaged) closed ribbon operators ${\cal R}_{\alpha}[G,H]$ discussed in section \ref{ribbonop}.  We remind the reader that the ribbon operators are associated to a closed path $\alpha$ and its shadow $\alpha'$, which needs to go along the graph $\Gamma$. The ribbon operator imposes the holonomy $G$ for $\alpha'$ via a factor $\delta(G,h_{\alpha'})$, with which the wave function is multiplied. Furthermore the holonomy of each link crossed by $\alpha$ is shifted from the left by $H^{-1}$ parallel transported along $\alpha'$ to some node on $\alpha'$. 

We saw that demanding that $G$ and $H$ commute we ensure that the ribbon operator does not violate any flatness constraint on the surface. We furthermore gauge averaged the ribbon operators to avoid a possible violation of gauge invariance. 

In addition we need to ensure that the ribbon operator leaves the flatness constraints $\{{\bf F}_t\}_{t\in \Delta}$ invariant. To this end we consider the following subclass of ribbon operators ${\cal R}_{\alpha}[G,H]$:  We choose $\alpha$ (and therefore $\alpha'$) to be isotopy equivalent to the curve $t\cap \Sigma(\Delta)$ for some triangle $t$ of the triangulation. (We will later consider more generalized operators.) To preserve the constraints $h_{\alpha'}=\mathbb{I}$, we will only consider the ribbon operators ${\cal R}_{\alpha}[G,H]$ with $G=\mathbb{I}$. The gauge averaging of such operators then simplifies and reduces the parameter $H$ to its conjugacy class $C$.  We denote the resulting operators by ${\cal R}_{t}[C]$.

The operators ${\cal R}_{t}[C]$ leave all the flatness constraints $\{{\bf F}_{t'}\}_{t'\in \Delta}$ invariant: It is clear that ${\cal R}_{t}[C]$ leaves the flatness constraint coming from $t$ itself invariant. Furthermore the closed curves $t \cap \Sigma(\Delta)$ and $t' \cap \Sigma(\Delta)$ for two different triangles $t$ and $t'$ do not intersect each other on $\Sigma(\Delta)$. Thus even if ${\cal R}_{t}[C]$ goes along a curve $\alpha$ that does intersect $t' \cap \Sigma(\Delta)$ it has to intersect it an even number of times as $\alpha$ needs to be isotopy equivalent to $t \cap \Sigma(\Delta)$ which does not intersect $t' \cap \Sigma(\Delta)$. Hence the holonomy associated to $t' \cap \Sigma(\Delta)$ (or to a curve isotopy equivalent to $t' \cap \Sigma(\Delta)$) will not be affected by the translational part of the action of ${\cal R}_{t}[C]$.

We can thus associated to each triangle $t$ ribbon operators ${\cal R}_{t}[C]$, labelled by a conjugacy class $C$ of ${\cal G}$. This ribbon operator changes the holonomies around the edges bounding the triangle $t$. This is the same action as for the (gauge averaged) integrated flux operator associated to a triangle $t$ and defined in \cite{DG14,BDG15}. 

We can also consider ribbon operators associated to more general curves than thus arising from one triangle $t$: For instance we can take two adjacent triangles $t,t'$ and consider their induced curves $t\cap \Sigma(\Delta)$ and $t'\cap \Sigma(\Delta)$.  We can then isotopically deform e.g. $t'\cap \Sigma(\Delta)$ to a curve $\beta$ such that $\beta$ agrees with $t\cap \Sigma(\Delta)$ on the part of the curve running on the tube ${\cal T}$ surrounding the edge shared by $t$ and $t'$. This defines a merging of the curves $t\cap \Sigma(\Delta)$ and $\beta$, denoted by $(t \circ t') \cap \Sigma(\Delta)$ and given by the set $( t\cap \Sigma(\Delta)) \cup \beta / (( t\cap \Sigma(\Delta)) \cap \beta)$.  

We can then consider a ribbon operator ${\cal R}_{t\circ t'}[C]$ associated to the curve $(t \circ t') \cap \Sigma(\Delta)$. This ribbon operator will shift the $h$--holonomies around the edges adjacent to the triangles $t$ and $t'$ but for the one edge shared by $t$ and $t'$ and along which we merged the two triangle curves to one curve. (Even if the two triangles have more than one edge in common it might not be possible to merge further parts of their curve, see the following discussion.)

This procedure of merging the curves induced by the triangles can be generalized to an arbitrary number of triangles. In this way we can consider ribbon operators associated to curves going around a number of triangles. Note however that it is {\it in general not} possible to merge the curves arising from triangles meeting at a vertex $v$, so that the merged curve does not visit ${\cal S}_v$ anymore. There might be punctures on   ${\cal S}_v$  (resulting from further edges adjacent to $v$) preventing such a merging. One can nevertheless consider a curve, e.g. resulting from the maximally possible merging of three triangles $t,t',t''$ meeting at a vertex. However even if the three triangles close around a vertex $v$ to a surface we might not be able to merge the three triangles along all the (three) shared edges. That is the merged path will have to include two parts along the same tube ${\cal T}$ surrounding one of the shared edges:  the path has to go on this tube ${\cal T}$ to ${\cal S}_v$, encircle some punctures on ${\cal S}_v$ and then come back again via the tube ${\cal T}$.  (We will discuss such a case in section \ref{4simplex}.)

Thus also the ribbon associated to such a merged path has to go back and forth along the tube ${\cal T}$. This can then have a non--trivial action on the $h$--holonomy going around the tube, as the translational action of the two pieces of the ribbon differ by the holonomies picked up by going around the punctures on ${\cal S}_v$.

In short we see that the ribbon operators associated to the merging of several triangle curves will not only depend on the triangles itself, but also on the details on how we merge the associated triangle curves. This determines along which path the group element $H$  is parallel transported, by which we translate each of the holonomies that are crossed by the ribbon.  

Note that this dependence does also appear for the integrated flux operators in $(3+1)$D defined in \cite{DG14,BDG15}. There the exponentiated flux operators are associated to a surface glued from triangles. But for each such surface we have also to specify a `surface tree' that describes the parallel transport of the translational parameter $H$ for each of the triangles of the surface. 

In both, the Heegaard surface and the 3D description, we can also consider a closed surface made out of triangles. However, for the reasons discussed above, in the ribbon case we will in general {\it not} be able to merge the curves induced by the triangles, so that the merged curve is equivalent to a contractible curve on $\Sigma(\Delta)$ and the associated ribbon therefore trivial. We will rather have a merged curve visiting some or (in general) all of the spheres ${\cal S}_v$ associated to the vertices $v$ adjacent to the triangles making up the glued surface. In general the closed surface made out of triangles is cut open by a connected and spanning tree made out of the edges of the surface. The merged curve runs along the tubes surrounding the edges of the tree and furthermore traverses these tubes twice in opposite directions.  The action of the ribbon associated to the merged curve is only non--trivial because of the difference in parallel transport for the two parts of the curve traversing each tube.

Also the exponentiated flux operators  associated to closed surfaces as defined in \cite{DG14,BDG15} have in general a non--trivial action due to the difference in parallel transport along some cut of the surface. This difference is only relevant if there is curvature, and, as the closed surface operators measure torsion, this effect has been named curvature induced torsion in \cite{DG14}.

\section{Example: Boundary of a 4--simplex}\label{4simplex}

As an example we consider the three--sphere $S^3$ triangulated by the boundary of a 4--simplex $\Delta=\sigma^4$.  We can identify $S^3$ with the  compactified space ${\mathbb R}^3$. This allows us to think of the boundary of the 4--simplex as embedded into ${\mathbb R}^3$. The one--skeleton of the triangulation agrees then with the one--skeleton obtained by subdividing one tetrahedron into four tetrahedra, see figure \ref{Fsimplex}. The four tetrahedra agree with four of the five tetrahedra of the 4--simplex, the fifth tetrahedron is given by the outside region of the subdivided tetrahedron (or by the complement of the subdivided tetrahededron in $S^3$).

We now have to consider a regular neighbourhood of the one--skeleton, that is blow--up the one--skeleton to a handlebody. We are interested in the boundary of this handlebody which defines a Heegaard surface $\Sigma(\Delta)$, depicted in figure \ref{Fsimplex}.

On this surface $\Sigma(\Delta)$ we have to choose a graph as described in section \ref{graph}.  As described there we can construct this graph separately on the (punctured) spheres and on the tubes of the Heegaard surfaces $\Sigma(\Delta)$. 

The tubes just carry one link of the graph that has to coincide with the part of a curve $t\cap \Sigma(\Delta)$ induced by one  of the triangles. For each edge of the triangulation, that is for each tube of the Heegaard surface, we have therefore to select one triangle $t$ adjacent to this edge. 
 We make the following choice: 
 For the edges of the subdivided tetrahedron we choose the links to go along the inside facing part of the tubes. That is for an edge $e(ij)$ with $i,j=1,\ldots,4$ we choose the triangle $t(ij5)$.    
 For an edges  $e(i5)$ with $i=1,2,3$ we choose the triangle  $t(i54)$. And for the edge $e(45)$ we choose the triangle $t(451)$.

\begin{figure}[!]
	\centering
	\begin{minipage}[b]{0.33\textwidth}
		\centering
		\includegraphics[scale = 0.9]{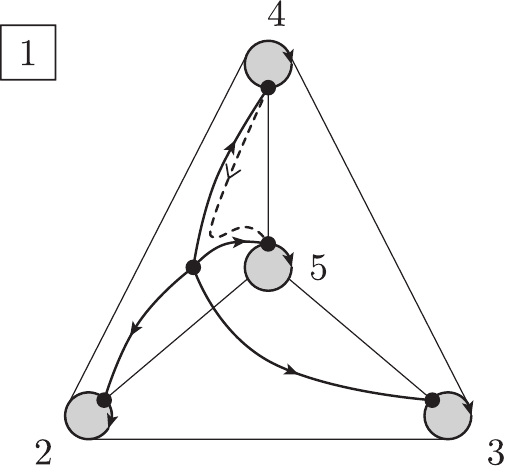}	
	\end{minipage}\q \q 
	\begin{minipage}[b]{0.33\textwidth}
	\centering
		\includegraphics[scale = 0.9]{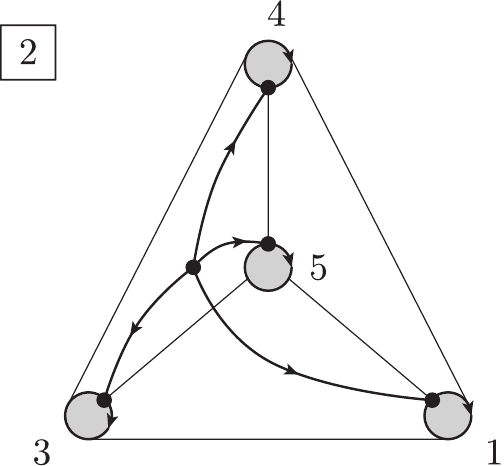}	
	\end{minipage}\\[4em]
	\begin{minipage}[b]{0.3\textwidth}
	\centering
		\includegraphics[scale = 0.9]{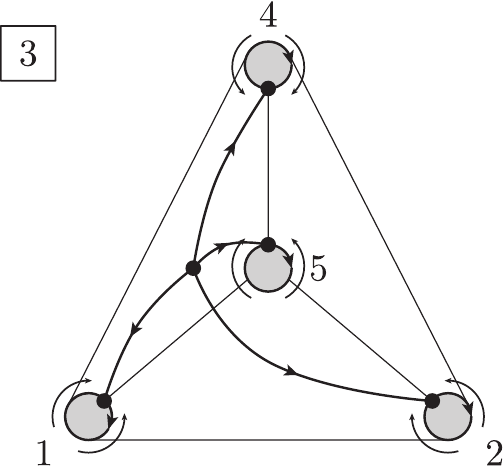}	
	\end{minipage}\q\q
	\begin{minipage}[b]{0.3\textwidth}
	\centering
		\includegraphics[scale = 0.9]{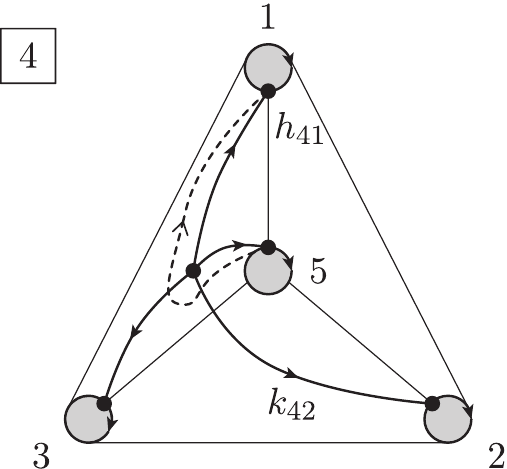}	
	\end{minipage}\q \q
	\begin{minipage}[b]{0.3\textwidth}
	\centering
		\includegraphics[scale = 0.9]{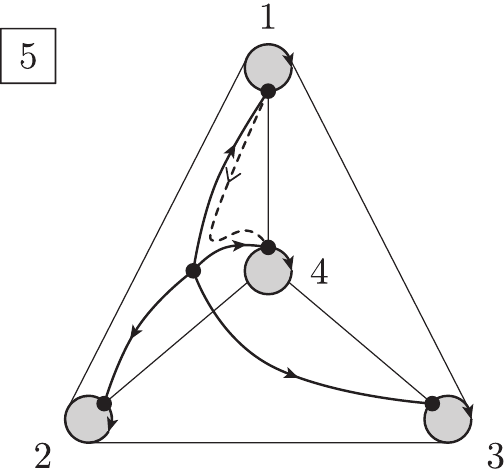}	
	\end{minipage}
	\caption{  This figure shows the five punctured spheres ${\cal S}_i,\,i=1,\ldots 5$ (in a planar representation) and the links of $\Gamma$	on these spheres.  The punctures are shown as grey disks. The spheres are glued to tubes, surrounding the edges of the triangulation, via the punctures. Thus each puncture on the sphere ${\cal S}_i$ can be labeled by the sphere ${\cal S}_j$ to which the glued tubes lead. The punctures are surrounded by links oriented clockwise as shown for the sphere ${\cal S}_4$. 
	The thin lines connecting the punctures show how the triangles cut through the punctured spheres and lead to curves on the Heegaard surface.  In the picture for ${\cal S}_3$ we have indicated how to deform the curves induced by the triangles so that these curves run along the links of the graph $\Gamma$. 
	The dashed line represents the path $\alpha$ defining the action of the ribbon operator $\mathcal{R}_{451}[H]$.  \label{Fcrit} }
\end{figure}

It is more involved to decide on a graph for the  punctured spheres around each vertex. But here we can actually draw  the punctured spheres (in a planar way) associated to each vertex, see figure \ref{Fcrit}.  In all cases we have four--punctured spheres. The four punctures are forming the four vertices of a flattened (on top view of a) tetrahedron. The edges of this (auxiliary) tetrahedron do indeed represent the curves resulting from the triangles of the triangulation cutting the punctured sphere in question. 

With our choice of links along the tubes connecting the spheres we have also chosen the `marked points' of the punctures, that is the point on the boundary of the puncture  at which the link coming from the tube will emerge.  For each punctured sphere we have to choose one node and four links connecting this node to the marked points on the punctures.  We present such a choice in figure \ref{Fcrit}.

There is one further choice to make, namely how to isotop the parts of the curves resulting from triangles cutting the Heegaard surfaces, which do not run already along the links of the graphs.  (This has to be done in particular for the part of the curves running along the tubes.) We indicate in figure  \ref{Fcrit}  in which way the triangle curves are isotoped around the punctures of a given sphere.

Let us consider the sphere ${\cal S}_1$ associated to the vertex $v_1$ in more detail. (As we have made all choices similar for the other spheres the same conclusions will hold there.)  The four punctures of the sphere can be labelled by the vertices which are connected (through tubes) to these punctures.  In trying to construct a path  along the graph which is isotopic to the curve induced by a given triangle we have to be careful in how the triangle curve, which emerges from one puncture and goes to another one, surrounds the remaining punctures.

For instance we see that  the curve given by  $t(214) \cap {\cal S}_1$ is isotopic to the following path along the graph $\Gamma$: we follow the link from puncture $p_4$ to the 4--valent `central' node $n_1$ and then follow the link to the marked point of the puncture $p_2$. This holds also for the triangles $t(213),t(415),t(215)$. That is for four out of six triangles we can just follow the canonical path along the graph. (Note that this holds only for the corners of the triangles at $v_1$, the same triangles will in general involve more complicated paths at other vertices.)  
To get however a path along the graph that is isotopic to  $t(413) \cap {\cal S}_1$,  we do however have to  -- in addition to the link from $p_4$ to $n_1$ and from $n_1$ to $p_3$ --  include a cycle around the puncture $p_5$. Also for the triangle $t(513)$ we have to include a cycle around the puncture $p_5$ itself.  

In fact for each of the five spheres we have four triangles which induce curves which are already isotopic to the canonical path along the graph. For the remaining two triangles we need however to include a cycle around a puncture, which is always the central puncture in the figures \ref{Fcrit}. 

~\\

Let us summarize our notation for the different kinds of holonomies along the Heegaard surface: For $i=1,\ldots, 5$ we define
\ba
k_{ij}  \q && \text{the holonomy from the node $n_i$ associated to the vertex $i$ to the puncture $p_j$,} \nn\\
g_{ij} \q && \text{the holonomy (along tube) from the (marked point on the) puncture $p_i$ on ${\cal S}_j$}\nn\\
 &&\text{to (the marked point on) the puncture $p_j$ on ${\cal S}_i$,} \nn\\
h_{ij} \q && \text{the holonomy cycling clockwise  the puncture $p_j$ on the sphere ${\cal S}_i$ associated to $v_i$}.\q
\ea

We can absorb the $k$ holonomies  into the $g$ and $h$ holonomies
\ba
\tilde g_{ij}  &=&  k^{-1}_{ij} g_{ij} k_{ji} \q ,\nn\\
\tilde h_{ij} &=&                           k_{ij}^{-1} h_{ij} k_{ij} \q .
\ea
This way the $\tilde g$ and $\tilde h$ holonomies start and end at the  `central' nodes $\{n_i\}_{i=1}^5$.

We can now give the holonomies along the curves induced by the triangles. With $t(kji)$ we denote the curve starting at vertex $i$ going to vertex $j$, then $k$ and then back to $i$. The flatness constraints induced by the triangles are as follows:
\ba\label{4.3}
&  t(231): \q \tilde g_{12} \tilde g_{23} \tilde g_{31} \,=\, \mathbb{I}  \q ,
\q \q  & t(125):\q  \tilde g_{25} \tilde g_{51} \tilde g_{12} \tilde h_{25} \,=\, \mathbb{I}  \q ,\nn\\
& t(514): \q \tilde g_{45} \tilde g_{51} \tilde g_{14} \,=\, \mathbb{I}  \q,
\q\q & t(234):\q \tilde g_{34} \tilde g_{42} \tilde g_{23} \tilde h_{35}  \,=\, \mathbb{I}  \q ,\nn\\
& \q t(524):    \q \tilde g_{45} \tilde g_{52} \tilde g_{24} \tilde h_{45}\,=\, \mathbb{I}  \q ,
\q\q & t(314): \q\tilde g_{14} \tilde g_{43} \tilde g_{31} \tilde h_{15}  \,=\, \mathbb{I}  \q ,\nn\\
& \q  t(354): \q \tilde g_{54} \tilde g_{43} \tilde g_{35} \tilde h_{54} \,=\, \mathbb{I}  \q ,
\q\q & t(315): \q \tilde g_{15}  \tilde h_{54}^{-1} \tilde g_{53} \tilde g_{31} \tilde h_{15} \,=\, \mathbb{I}  \q ,\nn\\
& \q t(235): \q \tilde g_{35} \tilde g_{52} \tilde g_{23} \tilde h_{35} \,=\, \mathbb{I}  \q ,
\q\q & t(124):\q \tilde g_{24} \tilde h_{45} \tilde g_{41}\tilde g_{12} \tilde h_{25}  \,=\, \mathbb{I}  \q . 
\ea

These conditions determine 6 variables $\tilde g_{ij}$ in terms of the $\tilde h_{ij}$ and the remaining four $\tilde g_{ij}$. Choosing these four variables to be $\tilde g_{12},\tilde g_{13},\tilde g_{14}$ and $\tilde g_{15}$ (corresponding to an allowed gauge fixing) we obtain
\ba
\tilde g_{23} &=& \tilde g_{21} \tilde g_{13} \label{4.4} \\
\tilde g_{45} &=& \tilde g_{41} \tilde g_{15}\label{4.5} \\
\tilde g_{43}&=& \tilde g_{41} (\tilde h_{15})^{-1} \tilde g_{13}  \label{4.6}\\
\tilde g_{24} &=& (\tilde h_{25})^{-1} \tilde g_{21} \tilde g_{14} (\tilde h_{45})^{-1}\label{4.7} \\
\tilde g_{25} &=& (\tilde h_{25})^{-1} \tilde g_{21} \tilde g_{15} \label{4.8}\\
\tilde g_{35}&=& (\tilde h_{35})^{-1} \tilde g_{31} \tilde g_{12} (\tilde h_{25})^{-1} \tilde g_{21} \tilde g_{15} \label{4.9} \q .
\ea
The remaining $4$ equations in  (\ref{4.3}) are redundant, for instance by leading to the Bianchi identity $\tilde h_{51} \tilde h_{52} \tilde h_{53}  \tilde h_{54}=\mathbb{I}$.

~\\

Let us now discuss the action of the ribbon operators associated to the triangles. We start with a simple case, the ribbon associated to the triangle $t(431)$.

~\\
\noindent ${\bf t(431)}$:\\
Consider ${\cal R}_{431}[C]$, the ribbon around the triangle $t(431)$ with base node $n_1$ so that we have a path $\alpha$
\be\label{p431}
n_1 \rightarrow n_3 \rightarrow n_4 \rightarrow n_1 \q . 
\ee
The associated triangle flatness constraints is 
\ba
\mathbb{I}\,=\, \tilde g_{14} \tilde g_{43} \tilde g_{31} \tilde h_{15}  \q .
\ea
One can follow the path (\ref{p431}) such that the links of the graph are always to the right of $\alpha$. This path will only cross $h_{ij}$--holonomies with $i,j \neq 5$. Thus all the triangle flatness constraints will be preserved.

More in detail, we can again express the shifts in terms of $\tilde h$ and $\tilde g$ variables. The ribbon ${\cal R}_{431}[C]$ leads to the shifts:
\begin{align}
\begin{cases}
\tilde h_{13}^{-1} \rightarrow ( \tilde g_{13} \tilde g_{34} \tilde g_{41}) \, H^{-1} \, ( \tilde g_{14} \tilde g_{43} \tilde g_{31}) \, \tilde h_{13}^{-1}  \\
  \tilde h_{31}\rightarrow (  \tilde g_{34} \tilde g_{41}) \, H^{-1} \, ( \tilde g_{14} \tilde g_{43} ) \, \tilde h_{31}   \\
 \tilde h^{-1}_{34} \rightarrow (  \tilde g_{34} \tilde g_{41}) \, H^{-1} \, ( \tilde g_{14} \tilde g_{43} ) \, \tilde h_{34}^{-1}  \\
 \tilde h_{43} \rightarrow (   \tilde g_{41}) \, H^{-1} \, ( \tilde g_{14}  ) \, \tilde h_{43}   \\
\tilde h_{41}^{-1} \rightarrow (   \tilde g_{41}) \, H^{-1} \, ( \tilde g_{14}  ) \, \tilde h_{41}^{-1}  \\
\tilde h_{14} \rightarrow H^{-1} \,\tilde h_{14} \q\q\q\q\q\q\q\q\q\q \; .
\end{cases}
\end{align}
To make the ribbon operator gauge invariant,  the group element $H$ is group averaged (by adjoint action), so that only the information on the conjugacy class $C$ remains.

~\\
The ribbons   ${\cal R}_{423}, {\cal R}_{124}$ and ${\cal R}_{213}$  function analogously. Next we discuss a case in which the ribbon also affects $\tilde h_{i5}$ holonomies and thus might a priori violate the triangle flatness constraints. 

~\\
\noindent ${\bf t(451)}$:\\
We consider the path $\alpha$:
\be\label{p451}
n_1 \rightarrow n_5 \rightarrow n_4 \rightarrow n_1 \q . 
\ee
We see that some  $\tilde h_{i5}$ holonomies are shifted but also the  $k_{42}$ and $k_{43}$ holonomies. Despite this we can again express everything in terms of $\tilde h$ and $\tilde g$ variables.

The crossing of the ribbon over links carrying $h$--holonomies leads to the following shifts: (We will make use of the triangle flatness constraint $\tilde g_{14} \tilde g_{45} \tilde g_{51} = \mathbb{I}$.)
\begin{align}
\begin{cases}
\tilde h^{-1}_{15}   \rightarrow H^{-1} \, \tilde h^{-1}_{15} \\
\tilde h_{51}  \rightarrow   \tilde g_{51}  \,H^{-1} \, \tilde g_{15} \, \tilde h_{51} \\
\tilde h^{-1}_{54} \rightarrow \tilde g_{51}   H^{-1} \tilde g_{15} \, \tilde h^{-1}_{54} \\
\tilde h_{45}  \rightarrow  \tilde g_{41} \, H^{-1} \tilde g_{14} \, \tilde h_{45} \\
\tilde h_{41}^{-1}  \rightarrow   \tilde g_{41} \, H^{-1} \tilde g_{14} \, \tilde h_{41}^{-1} \\
\tilde h_{14}   \rightarrow   H^{-1}\, \tilde h_{14}  \q\q\q\q\q\q .
\end{cases}
\end{align}

Furthermore the crossing of the $k_{42}$ and $k_{43}$ variables influences the following variables:
\begin{align}
\begin{cases}
\tilde g_{24} \rightarrow \tilde g_{24}\, \tilde g_{41} H \tilde g_{14} \\
\tilde h_{42}  \rightarrow \tilde g_{41} H^{-1} \tilde g_{14} \tilde h_{42} \tilde g_{41} H \tilde g_{14} \\
\tilde g_{34} \rightarrow \tilde  g_{34} \tilde g_{41} H \tilde g_{14} \\
\tilde h_{43}  \rightarrow \tilde g_{41} H^{-1} \tilde g_{14} \tilde h_{43} \tilde g_{41} H \tilde g_{14} \q .
\end{cases}
\end{align}

The shifts do affect a priori the triangle flatness constraints (\ref{4.6}) and (\ref{4.7}). One can check however that the shifts of the various holonomies involved cancel out, and thus the triangle flatness constraints remain invariant.

Again the ribbons associated to the triangles $t(524), t(512), t(523)$ and $t(354)$ work similarly.  As the last slightly more subtle case we discuss the ribbon associated to the triangle $t(513)$.

~\\
\noindent ${\bf t(513)}$:\\
Let us now consider the ribbon operator associated to the triangle $t(135)$ with based node $n_5$ so that the path $\alpha$ is given by 
\be
	n_5 \rightarrow n_3 \rightarrow n_1 \rightarrow n_5 \q . 
\ee
The corresponding triangle flatness constraint reads
\be
	\tilde{g}_{51}\tilde{h}^{-1}_{15}\tilde{g}_{13}\tilde{g}_{35}\tilde{h}_{54} =\mathbb{I} \q . 
\ee
This example is more cumbersome than the previous ones since the ribbon crosses the link $h_{15}$ which is part of the shadow path $\alpha'$. This is the situation described in appendix A. In order to circumvent this difficulty, we introduce additional links and decorate them with auxiliary holonomies. Doing so we must ensure that the flatness constraint as well as the gauge invariance are still satisfied and therefore the number of degrees of freedom is preserved. This refining of the graph is performed around the puncture $p_5$ and can be graphically represented as follows
\begin{align}
	\begin{array}{c}\includegraphics[scale = 1]{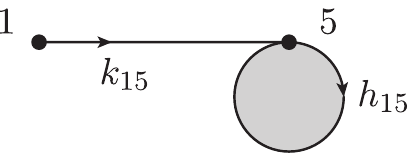}\end{array} \q \longrightarrow \q
	\begin{array}{c}\includegraphics[scale = 1]{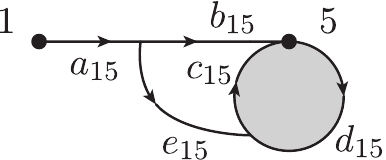}\end{array} .
\end{align}
Furthermore, the enforcement of the constraints impose the following expressions between the original variables and the auxiliary ones
\be
	k_{15} = b_{15}a_{15} \q , \q h_{15} = c_{15} d_{15}\q , \q e_{15}c_{15}b_{15}^{-1} = \mathbb{I}\q .
	\label{corres}
\ee
It is now possible to define a ribbon which does not cross any holonomy appearing in the definition of the path $\alpha'$. Indeed, let us for instance consider the following ribbon
\begin{align}
	\begin{array}{c}\includegraphics[scale = 1]{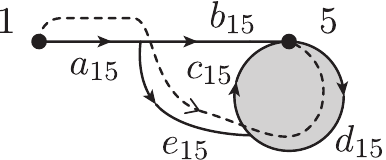}\end{array}
\end{align}
which crosses the links decorated by the holonomies $b_{15}$ and $c_{15}$ whereas the corresponding path $\alpha'$ is associated to the holonomy $d_{15}^{-1}e_{15}a_{15}$. Putting everything together, we obtain that the ribbon operator produces the following shifts: 
\begin{align}
	\begin{cases}
		\tilde{h}_{53}^{-1} \rightarrow (\tilde{g}_{53}\tilde{g}_{31}\tilde{h}_{15}\tilde{g}_{15})H^{-1}
		(\tilde{g}_{51}\tilde{h}_{15}^{-1}\tilde{g}_{13}\tilde{g}_{35})\tilde{h}_{53}^{-1} \\[0.4em]
		\tilde{h}_{35} \rightarrow (\tilde{g}_{31}\tilde{h}_{15}\tilde{g}_{15})H^{-1}
		(\tilde{g}_{51}\tilde{h}_{15}^{-1}\tilde{g}_{13})\tilde{h}_{35} \\[0.4em]
		{k}_{32}^{-1} \rightarrow (\tilde{g}_{31}\tilde{h}_{15}\tilde{g}_{15})H^{-1}
		(\tilde{g}_{51}\tilde{h}_{15}^{-1}\tilde{g}_{13}){k}_{32}^{-1} \\[0.4em]
		\tilde{h}_{31}^{-1} \rightarrow (\tilde{g}_{31}\tilde{h}_{15}\tilde{g}_{15})H^{-1}
		(\tilde{g}_{51}\tilde{h}_{15}^{-1}\tilde{g}_{13})\tilde{h}_{31}^{-1} \\[0.4em]
		\tilde{h}_{13} \rightarrow (\tilde{h}_{15}\tilde{g}_{15})H^{-1}
		(\tilde{g}_{51}\tilde{h}_{15}^{-1})\tilde{h}_{13} \\[0.4em]
		{k}_{12}^{-1} \rightarrow (\tilde{h}_{15}\tilde{g}_{15})H^{-1}
		(\tilde{g}_{51}\tilde{h}_{15}^{-1}){k}_{12}^{-1} \\[0.4em]
		{k}_{14}^{-1} \rightarrow (\tilde{h}_{15}\tilde{g}_{15})H^{-1}
		(\tilde{g}_{51}\tilde{h}_{15}^{-1})\tilde{k}_{14}^{-1} \\[0.4em]
		\tilde{h}_{51} \rightarrow H^{-1}\tilde{h}_{51}
	\end{cases}
\end{align}
and
\begin{align}
	\begin{cases}
		k_{15}\rightarrow h_{15}g_{15}k_{51}H \tilde{g}_{51}\tilde{h}_{15}^{-1} \\[0.4em]
		h_{15} \rightarrow h_{15}g_{15}k_{51}H k_{51}^{-1}g_{51}
	\end{cases}
\end{align}
where we have used for these last two shifts the results from Appendix A.
The shifts of the $k$-holonomies $k_{32}, k_{12}, k_{14}$ and $k_{15}$ and the $h$-holonomy $h_{15}$ presented above influence the following variables:
\begin{align}
	\begin{cases}
		\tilde{g}_{32} \rightarrow (\tilde{g}_{31}\tilde{h}_{15}\tilde{g}_{15})H^{-1}(\tilde{g}_{51}\tilde{h}_{15}^{-1}\tilde{g}_{13})\tilde{g}_{32} \\[0.4em]
		\tilde{h}_{32} \rightarrow (\tilde{g}_{31}\tilde{h}_{15}\tilde{g}_{15})H^{-1}(\tilde{g}_{51}\tilde{h}_{15}^{-1}\tilde{g}_{13})\tilde{h}_{32} 
		(\tilde{g}_{31}\tilde{h}_{15}\tilde{g}_{15})H(\tilde{g}_{51}\tilde{h}_{15}^{-1}\tilde{g}_{13}) \\[0.4em]
		\tilde{g}_{12} \rightarrow (\tilde{h}_{15}\tilde{g}_{15})H^{-1}(\tilde{g}_{51}\tilde{h}_{15}^{-1})\tilde{g}_{12} \\[0.4em]
		\tilde{h}_{12} \rightarrow  (\tilde{h}_{15}\tilde{g}_{15})H^{-1}(\tilde{g}_{51}\tilde{h}_{15}^{-1})\tilde{h}_{12}
		 (\tilde{h}_{15}\tilde{g}_{15})H(\tilde{g}_{51}\tilde{h}_{15}^{-1}) \\[0.4em]
		\tilde{g}_{14} \rightarrow (\tilde{h}_{15}\tilde{g}_{15})H^{-1}(\tilde{g}_{51}\tilde{h}_{15}^{-1})\tilde{g}_{14} \\[0.4em]
		\tilde{h}_{14} \rightarrow  (\tilde{h}_{15}\tilde{g}_{15})H^{-1}(\tilde{g}_{51}\tilde{h}_{15}^{-1})\tilde{h}_{14}
		 (\tilde{h}_{15}\tilde{g}_{15})H(\tilde{g}_{51}\tilde{h}_{15}^{-1}) \\[0.4em]
		\tilde{g}_{15}  \rightarrow 
		(\tilde{h}_{15}\tilde{g}_{15})H^{-1}(\tilde{g}_{51}\tilde{h}_{15}^{-1})\tilde{g}_{15} \\[0.4em]
		\tilde{h}_{15} \rightarrow \tilde{h}_{15}\tilde{h}_{15}\tilde{g}_{15}H\tilde{g}_{51}\tilde{h}_{15}^{-1}
	\end{cases}
\end{align}
Despite these additional shifts, the flatness constraints are not violated. Indeed, the different shifts cancel each other such that all the flatness constraints presented earlier remain invariant. In particular, we have the following trivial transformations
\begin{align}
	\begin{cases}
		\tilde{g}_{12}\tilde{g}_{23}\tilde{g}_{31}  = \mathbb{I}    \rightarrow \tilde{h}_{15}\tilde{g}_{15}H^{-1}\tilde{g}_{51}\tilde{h}_{15}^{-1}\tilde{g}_{12}\tilde{g}_{23}\tilde{g}_{31}\tilde{h}_{15}\tilde{g}_{15}H\tilde{g}_{51}\tilde{h}_{15}^{-1}\tilde{g}_{13}\tilde{g}_{31} = \mathbb{I} \\[0.4em]
		\tilde{g}_{35}\tilde{g}_{52}\tilde{g}_{23}\tilde{h}_{35} = \mathbb{I}\rightarrow \tilde{g}_{35}\tilde{g}_{52}\tilde{g}_{23}\tilde{g}_{31}\tilde{h}_{15}\tilde{g}_{15}H\tilde{g}_{51}\tilde{h}_{15}^{-1}\tilde{g}_{13}
	  \tilde{g}_{31}\tilde{h}_{15}\tilde{g}_{15}H^{-1}
		\tilde{g}_{51}\tilde{h}_{15}^{-1}\tilde{g}_{13}\tilde{h}_{35}= \mathbb{I} \\[0.4em]
		\tilde{g}_{15}\tilde{h}_{54}^{-1}\tilde{g}_{53}\tilde{g}_{31}\tilde{h}_{15} = \mathbb{I} \rightarrow \tilde{h}_{15}\tilde{g}_{15}H^{-1}\tilde{g}_{51}\tilde{h}_{15}^{-1}\tilde{g}_{15} \tilde{h}_{54}^{-1}\tilde{g}_{53}\tilde{g}_{31}\tilde{h}_{15}\tilde{h}_{15}\tilde{g}_{15}H\tilde{g}_{51}\tilde{h}_{15}^{-1}= \mathbb{I}
	\end{cases}
\end{align}
and one can check the remaining ones similarly. \\
\newpage
~\\
{\bf {Merging of two triangles:}}\\
We can now consider the case of a ribbon associated with the merging of two triangles. Let us for instance consider the triangles $t(431)$ and $t(412)$ which have in common the tube going from the puncture $1$ to the puncture $4$. We are looking for a ribbon operator $\mathcal{R}_{4312}[H]$ associated to the curve going around the two triangles so that we have the following path $\alpha$
\be
 n_2 \rightarrow n_1 \rightarrow n_3 \rightarrow n_4 \rightarrow n_2 \q . 
\ee
The flatness constraints for the triangle $t(431)$ and $t(412)$ are respectively given by $\tilde{g}_{14}\tilde{g}_{43}\tilde{g}_{31}\tilde{h}_{15} = \mathbb{I}$ and $\tilde{g}_{24}\tilde{h}_{45}\tilde{g}_{41}\tilde{g}_{12}\tilde{h}_{25} = \mathbb{I}$ so that the flatness constraint assiociated with the merging of the two triangles reads
\be
	\tilde{g}_{24}\tilde{h}_{45}\tilde{g}_{43}\tilde{g}_{31}\tilde{h}_{15}\tilde{g}_{12}\tilde{h}_{25} = \mathbb{I} \q.
\ee
The action of the ribbon then produces the following shifts of holonomies
\begin{align}
	\begin{cases}
		\tilde{h}_{21}^{-1} \rightarrow
		(\tilde{g}_{21}\tilde{h}_{15}^{-1}\tilde{g}_{13}\tilde{g}_{34}\tilde{h}_{45}^{-1}\tilde{g}_{42})H^{-1}
		(\tilde{g}_{24}\tilde{h}_{45}\tilde{g}_{43}\tilde{g}_{31}\tilde{h}_{15}\tilde{g}_{12})\tilde{h}^{-1}_{21} \\[0.4em]
		\tilde{h}_{12} \rightarrow
		(\tilde{h}_{15}^{-1}\tilde{g}_{13}\tilde{g}_{34}\tilde{h}_{45}^{-1}\tilde{g}_{42})H^{-1}
		(\tilde{g}_{24}\tilde{h}_{45}\tilde{g}_{43}\tilde{g}_{31}\tilde{h}_{15})\tilde{h}_{12} \\[0.4em]
		{k}_{14}^{-1} \rightarrow
		(\tilde{h}_{15}^{-1}\tilde{g}_{13}\tilde{g}_{34}\tilde{h}_{45}^{-1}\tilde{g}_{42})H^{-1}
		(\tilde{g}_{24}\tilde{h}_{45}\tilde{g}_{43}\tilde{g}_{31}\tilde{h}_{15}){k}_{14}^{-1} \\[0.4em]
		\tilde{h}_{13}^{-1} \rightarrow
		(\tilde{g}_{13}\tilde{g}_{34}\tilde{h}_{45}^{-1}\tilde{g}_{42})H^{-1}
		(\tilde{g}_{24}\tilde{h}_{45}\tilde{g}_{43}\tilde{g}_{31})\tilde{h}_{13}^{-1} \\[0.4em]
		\tilde{h}_{31} \rightarrow
		(\tilde{g}_{34}\tilde{h}_{45}^{-1}\tilde{g}_{42})H^{-1}
		(\tilde{g}_{24}\tilde{h}_{45}\tilde{g}_{43})\tilde{h}_{31} \\[0.4em]
		\tilde{h}_{34}^{-1} \rightarrow
		(\tilde{g}_{34}\tilde{h}_{45}^{-1}\tilde{g}_{42})H^{-1}
		(\tilde{g}_{24}\tilde{h}_{45}\tilde{g}_{43})\tilde{h}_{34}^{-1} \\[0.4em]
		\tilde{h}_{43} \rightarrow (\tilde{h}_{45}^{-1}\tilde{g}_{42})H^{-1}(\tilde{g}_{24}\tilde{h}_{45})\tilde{h}_{43} \\[0.4em]
		{k}_{41}^{-1} \rightarrow (\tilde{h}_{45}^{-1}\tilde{g}_{42})H^{-1}(\tilde{g}_{24}\tilde{h}_{45}){k}_{41}^{-1} \\[0.4em]
		\tilde{h}_{42}^{-1} \rightarrow \tilde{g}_{42}H^{-1}\tilde{g}_{24}\tilde{h}_{42}^{-1} \\[0.4em]
		\tilde{h}_{24} \rightarrow H^{-1}\tilde{h}_{24}
	\end{cases}
\end{align}
The only remarkable feature is the fact that the ribbon operator acts on both $k_{14}^{-1}$ and $k_{41}^{-1}$ which is not the case when considering the independent actions on triangles $t(431)$ and $t(412)$. For the action of the ribbon on $\tilde{g}_{14}$ both transformation compensate so that  $\tilde{g}_{14}$ remains unchanged. Indeed, we have
\be
	\tilde{g}_{14} \rightarrow
	(\tilde{h}_{15}^{-1}\tilde{g}_{13}\tilde{g}_{34}\tilde{h}_{45}^{-1}\tilde{g}_{42})H^{-1}
		(\tilde{g}_{24}\tilde{h}_{45}\tilde{g}_{43}\tilde{g}_{31}\tilde{h}_{15})\tilde{g}_{14}(\tilde{h}_{45}^{-1}\tilde{g}_{42})H(\tilde{g}_{24}\tilde{h}_{45}) = \tilde{g}_{14}
\ee
where we have used twice the flatness constraint on the triangle $t(413)$. Also $h_{14}$ and $h_{41}$ are not changed by the ribbon operator. We do have however a change of $\tilde h_{14}= k_{14}^{-1} h_{14} k_{14}$ and $\tilde h_{41} = k_{41}^{-1} h_{41} k_{41}$ due to the shift of $k_{14}$ and $k_{41}$. Note however that this change is by an adjoint action, so the (gauge invariant) conjugacy classes of $\tilde h_{14}$ and $\tilde h_{41}$ do not change.

Therefore, as previously discussed in the general case, the ribbon operator changes the holonomies associated with the edges adjacent to the triangles but for the edge shared by the triangles. Furthermore, since the remaining shifts are only about $h$-holonomies which do not influence the flatness constraints, we can confirm that the ribbon operator for the merging of two triangles as defined here is consistent.\\
 \newpage
~\\
{\bf {Merging of three triangles:}}\\
In the previous section, we discussed the possibility of merging the curves induced by more than two triangles. We consider here the situation of three triangles closing around a vertex $v$. For instance, we can consider the merging of the triangles $t(314), t(124)$ and $t(234)$ which close around the vertex $4$. We can do the merging in two ways: by deforming the triangle curves to the `outside' facing part of the tubes in figure \ref{Fsimplex}. The triangle curves can then indeed deformed and merged with each other across the `top' of the sphere ${\cal S}_4$ (referring again to figure \ref{Fsimplex}). The resulting path would then just go along the tubes connecting the vertices $v_1,v_2$ and $v_3$. 

Another possibility, which we will discuss here, as it is more generic, is to attempt to merge the three triangles by deforming the triangle curves to the `inside' facing parts of the tubes. In attempting to merge the curves of the three triangles across the (`inside' facing part of the) sphere $\mathcal{S}_4$ we encounter an obstacle, namely the tube ${\cal T}_{54}$. We therefore choose the merged path to 
go back and forth along the tube $\mathcal{T}_{34}$ so that it can go around the tube  ${\cal T}_{54}$. The full path $\alpha$ is given by 
\be
	n_2 \rightarrow n_1 \rightarrow n_3 \rightarrow n_4 \rightarrow n_3 \rightarrow n_2 \q . 
\ee
and we denote by $\mathcal{R}_{34312}[H]$ the ribbon associated to it. Furthermore, the flatness constraints associated with the triangles $t(314), t(124)$ and $t(234)$ respectively read $\tilde{g}_{14}\tilde{g}_{43}\tilde{g}_{31}\tilde{h}_{15} = \mathbb{I}$, $\tilde{g}_{24}\tilde{h}_{45}\tilde{g}_{41}\tilde{g}_{12}\tilde{h}_{25} = \mathbb{I}$ and $\tilde{g}_{34}\tilde{g}_{42}\tilde{g}_{23}\tilde{h}_{35} = \mathbb{I}$ so that the flatness constraint corresponding to the merged path is
\be
	\tilde{g}_{23}\tilde{h}_{35}\tilde{g}_{34}\tilde{h}_{45}\tilde{g}_{43}\tilde{g}_{31}\tilde{h}_{15}\tilde{g}_{12}\tilde{h}_{25} = \mathbb{I} \q .
	\label{flat3tri}
\ee
The shifts induced by the ribbon $\mathcal{R}_{34312}[H]$ are listed below:
\begin{align}
	\begin{cases}
	\tilde{h}_{21}^{-1} \rightarrow \tilde{h}_{25}H^{-1}\tilde{h}_{25}^{-1}\tilde{h}_{21}^{-1} \\[0.4em]
	\tilde{h}_{12} \rightarrow (\tilde{g}_{12}\tilde{h}_{25})H^{-1}(\tilde{h}_{25}^{-1}\tilde{g}_{21})\tilde{h}_{12}  \\[0.4em]
	k_{14}^{-1 }\rightarrow (\tilde{g}_{12}\tilde{h}_{25})H^{-1}(\tilde{h}_{25}^{-1}\tilde{g}_{21}) k_{14}^{-1} \\[0.4em]
	\tilde{h}_{13}^{-1} \rightarrow (\tilde{h}_{15}\tilde{g}_{12}\tilde{h}_{25})H^{-1}(\tilde{h}_{25}^{-1}\tilde{g}_{21}\tilde{h}_{15}^{-1})\tilde{h}_{13}^{-1} \\[0.4em]
	\tilde{h}_{31} \rightarrow (\tilde{g}_{34}\tilde{h}_{45}^{-1}\tilde{g}_{43}\tilde{h}_{35}^{-1}\tilde{g}_{32})H^{-1}
	(\tilde{g}_{23}\tilde{h}_{35}\tilde{g}_{34}\tilde{h}_{45}\tilde{g}_{43})\tilde{h}_{31} \\[0.4em]
	\tilde{h}_{34}^{-1} \rightarrow (\tilde{g}_{34}\tilde{h}_{45}^{-1}\tilde{g}_{43}\tilde{h}_{35}^{-1}\tilde{g}_{32})H^{-1}
	(\tilde{g}_{23}\tilde{h}_{35}\tilde{g}_{34}\tilde{h}_{45}\tilde{g}_{43})\tilde{h}_{34}^{-1} \\[0.4em]
	\tilde{h}_{43} \rightarrow (\tilde{h}_{45}^{-1}\tilde{g}_{43}\tilde{h}_{35}^{-1}\tilde{g}_{32})H^{-1}
	(\tilde{g}_{23}\tilde{h}_{35}\tilde{g}_{34}\tilde{h}_{45})\tilde{h}_{43} \\[0.4em]
	{k}_{41}^{-1} \rightarrow (\tilde{h}_{45}^{-1}\tilde{g}_{43}\tilde{h}_{35}^{-1}\tilde{g}_{32})H^{-1}
	\end{cases} , 
	\begin{cases}
	(\tilde{g}_{23}\tilde{h}_{35}\tilde{g}_{34}\tilde{h}_{45}){k}_{41}^{-1} \\[0.4em]
	{k}_{42}^{-1} \rightarrow (\tilde{g}_{43}\tilde{h}_{35}^{-1}\tilde{g}_{32})H^{-1}
	(\tilde{g}_{23}\tilde{h}_{35}\tilde{g}_{34}){k}_{42}^{-1} \\[0.4em]
	\tilde{h}_{43}^{-1} \rightarrow (\tilde{g}_{43}\tilde{h}_{35}^{-1}\tilde{g}_{32})H^{-1}
	(\tilde{g}_{23}\tilde{h}_{35}\tilde{g}_{34})\tilde{h}_{43}^{-1} \\[0.4em]
	\tilde{h}_{34} \rightarrow (\tilde{h}_{35}^{-1}\tilde{g}_{32})H^{-1}(\tilde{g}_{23}\tilde{h}_{35})\tilde{h}_{34} \\[0.4em]
	\tilde{h}_{32}^{-1} \rightarrow \tilde{g}_{32}H^{-1}\tilde{g}_{23}\tilde{h}_{32}^{-1} \\[0.4em]	
	\tilde{h}_{23} \rightarrow H^{-1}\tilde{h}_{23} \\[0.4em]
	{k}_{24}^{-1} \rightarrow H^{-1}k_{24}^{-1} \\[0.4em]
	\end{cases}
\end{align}
where we have used the flatness constraint (\ref{flat3tri}) to simplify the expression of the first shifts. First of all, we notice a shift for the variables $k_{14}^{-1},k_{41}^{-1}$ and $k_{42}^{-1},k_{24}^{-1}$ which are respectively associated with the edges shared by the triangles $t(314),t(124)$ and $t(124),t(234)$. As for the case with two triangles, these shifts are such that they compensate each other for the action on the variables $\tilde{g}_{14}$ and $\tilde{g}_{42}$:
\begin{align}
	\tilde{g}_{14} &\rightarrow
	\tilde{g}_{12}\tilde{h}_{25}H^{-1}\tilde{h}_{25}^{-1}\tilde{g}_{21}\tilde{g}_{14}\tilde{h}_{45}^{-1}\tilde{g}_{42}
	H\tilde{g}_{24}\tilde{h}_{45} = \tilde{g}_{12}\tilde{h}_{25}\tilde{g}_{24}\tilde{h}_{45} = \tilde{g}_{14} \\[0.4em]
	\tilde{g}_{42} &\rightarrow \tilde{g}_{43}\tilde{h}_{35}^{-1}\tilde{g}_{32}H^{-1}\tilde{g}_{23}\tilde{h}_{35}\tilde{g}_{34}\tilde{g}_{42}H = \tilde{g}_{42}
\end{align}
where we have again used the triangle flatness constraints. Also the $h$-holonomies $h_{14}$ and ${h}_{42}$ remain invariant. But again $\tilde h_{14},\tilde h_{41}$ and $\tilde h_{42},\tilde h_{24}$ are affected (by an adjoint action) due to the shift of $k_{14},k_{41}$ and $k_{42},k_{24}$, however the conjugacy classes remain invariant.

 Therefore, as before, we have holonomies carried by edges shared by two triangles which are left invariant under the action of the ribbon operator. However, since the path goes back and forth along the tube $\mathcal{T}_{34}$, there is now a non-trivial action on the holonomy going around this tube. Indeed, the holonomy is shifted twice so that the overall transformation becomes
\begin{align}
	\tilde{h}_{43} \rightarrow \q &
	(\tilde{h}_{45}^{-1}\tilde{g}_{43}\tilde{h}_{35}^{-1}\tilde{g}_{32})H^{-1}(\tilde{g}_{23}\tilde{h}_{35}\tilde{g}_{34}\tilde{h}_{45})\,\,\tilde{h}_{43} \,\,
	(\tilde{g}_{43}\tilde{h}_{35}^{-1}\tilde{g}_{32})H(\tilde{g}_{23}\tilde{h}_{35}\tilde{g}_{34}) \nn\\
	&=(\tilde{h}_{45}^{-1}\tilde{g}_{42} )H^{-1}( \tilde{g}_{24}\tilde{h}_{45})\,\,\tilde{h}_{43}\,\,
	\tilde{g}_{42}H\tilde{g}_{24}
\end{align}
This change in $\tilde h_{43}$  is due to the double crossing of the $h_{43}$ holonomies by the ribbons (and not due to a shift of the $k_{43}$ holonomy, which in fact remains invariant).  Moreover if $\tilde h_{45} \neq \mathbb{I}$ the change will in general also affect the conjugacy class of $\tilde h_{43}$.

\section{Generalization: including torsion degrees of freedom}\label{torsion}

We discussed here the Hilbert space of gauge invariant wave functions on the space of locally flat connections on ${\cal M}/\Delta_1$ and gauge invariant operators generating curvature excitations concentrated on $\Delta_1$.  Apart from curvature excitations one can also consider torsion excitations, that is violations of gauge invariance for the wave functions. With torsion excitations we describe any violation of gauge invariance of the wave functions. This comes from a geometric interpretation of the gauge theory variables. In an electro-magnetic interpretation these excitations correspond to electric charges. In fact, if one wishes to work with a (compact) Lie group ${\cal G}$  as in \cite{DG14a,DG14,BDG15}, instead of a finite group,  it is more convenient to introduce one root (a node of the graph), at which gauge invariance can be violated. 

In general, to allow for torsion degrees of freedom, it is convenient to add one link $l_v$ to each sphere $S_v$, that starts at the node $n_v$ and (without intersecting any other links of the graph) ends in a one--valent node $n_v^o$ on $S_v$.
One then demands that the wave functions are gauge invariant at all higher than one--valent nodes.

One can then still consider the gauge invariant operators we have discussed so far. In addition we have now also the options of considering not fully gauge invariant operators, that can generate torsion. Firstly, we can have open Wilson line operators that start and end at the nodes $n_v^o$. Furthermore we can also consider ribbon operators that still have to be associated to a closed curve $\alpha$,  but now with a start and end node $n_v^o$ of the shadow curve $\alpha'$.  In this case we do not need to group average (with the adjoint action) over the translation parameter $H$ anymore. This then allows also for more possibilities to combine the closed ribbon operators based on the same node $n_v^o$, as the translation parameter is then always transported to the same frame (namely the frame at $n_v^o$).

However, introducing the one--valent nodes at which gauge invariance of the wave functions might be violated we loose part of the freedom to deform the curves $\alpha$ (and thus $\alpha'$) to which the closed ribbons are associated. Basically we have to treat the one--valent nodes as punctures across which we cannot deform the curve $\alpha$ anymore.

 In the case that one just has one root node $n_r=n^o_v$ for a particular vertex $v$, one can choose for any closed ribbon a parallel transport path connecting the start and end point of $\alpha'$ with the root $n_r$. Thus all ribbon operators (and also all Wilson loops) can be based at this root $n_r$.  As described in \cite{BDG15}, this is a convenient set--up in order to have as a structure group ${\cal G}$ a proper Lie group.


\section{Discussion and Outlook}\label{outlook}

In this work we explained how the Hilbert space and operators for a $(2+1)$ dimensional theory of flat connections lead to a Hilbert space and operators for a $(3+1)$ dimensional theory of flat connections with curvature defects. A crucial point is to use the Heegaard surface that arises from the Heegaard splitting of the 3D manifold, describing the equal time hypersurface of a $(3+1)$ dimensional manifold. The Heegaard splitting can be based on a triangulation (or other polyhedral lattice), the curvature defects are then confined to the one--skeleton of this triangulation.  The  theory of flat connections on the 3D manifold can then be described in terms of the theory of flat connections on the 2D Heegaard surface, but equipped with additional flatness constraints. In particular we can express operators generating curvature defects for the 3D theory as (ribbon) operators acting on the space of flat connections on the 2D surface, satisfying the additional flatness constraints.

This presents an interesting example where a $(2+1)$D dimensional topological quantum field theory can be used to construct a Hilbert space, with a (triangulation independent) vacuum state and excitations for a $(3+1)$D dimensional theory. We believe that this technique can be applied to a wide range of $(2+1)$D dimensional TQFT's and thus would allow the construction and understanding of a wide range of $(3+1)$ dimensional TQFT's with defects.

Future work will focus on generalization of this technique to the Turaev Viro TQFT \cite{TV}, see \cite{BD2017}. This would then generalize the recent work \cite{DG16} from $(2+1)$ to $(3+1)$ dimensions, and provide us with a new ($(3+1)$D) realization of quantum geometry, in which the Hilbert spaces (based on a fixed triangulation) are finite dimensional. The geometrical interpretation of this representation has to be explored. We conjecture that, as in the $(2+1)$ dimensional case, states are peaked almost everywhere on homogeneously curved geometry. As such this new representation would incorporate a cosmological constant and realize a homogeneously curved version of Regge calculus constructed in \cite{Improved,NewRegge}.  Furthermore, techniques recently developed in \cite{BB} might allow further generalizations of this $(3+1)$  dimensional quantum geometry realization, such that the underlying vacuum is based on a TQFT which can distinguish four--dimensional topologies (or in condensed matter terminology can have degenerate ground states), see \cite{BD2017} for further discussion.

There is an interesting connection to recent work by Haggard, Han, Kaminski and Riello \cite{Aldo1,Aldo2,Aldo3,Aldo4}, which constructs  spin foam amplitudes for quantum gravity incorporating  a cosmological constant. This work also studies the semi--classical limit of the simplex amplitudes proposed in \cite{Aldo1,Aldo2,Aldo3,Aldo4} and the related (classical) phase space associated to the (boundary of a) 4-simplex. The resulting (graph) structures are very similar to our example in section \ref{4simplex}. However Haggard {\it et al.}  use a dual picture to ours, that is the defects are not given by the triangulation but by the dual graph.

 These curvature defects on the dual graph links break the flatness condition around the triangles and are interpreted as homogeneous curvature on the triangles dual to the links. This interpretation arises in an semiclassical analysis of their spin foam amplitude, which is peaked on holonomies that can be interpreted as coming from an homogeneously curved (Lorentzian and four-dimensional) simplicial geometry.
This hints to the fact that their classical description given in \cite{Aldo1,Aldo2,Aldo3,Aldo4} could arise as an interesting deformation of the classical limit of the
construction given in this paper. The question is then whether an application of the technique presented here to the Turaev--Viro model is in fact a quantization of the phase space described in \cite{Aldo1,Aldo2,Aldo3,Aldo4}.

More generally the $(2+1)$D (Levin-Wen) string nets \cite{LevinWen,Lan,Kir} encompass BF theory with finite groups and quantum groups at root of unity (or Turaev--Viro models) as well as models based on other types of fusion categories. It will be interesting to see whether, using the Heegaard splitting, all these models can be lifted to $(3+1)$ dimensions, and to study their excitation structure. In \cite{LevinWen, WalkerWang} Hamitlonians (or Hamiltonian constraints) have been suggested  that would realize a $(3+1)$ dimensional generalization of the Levin-Wen models. The techniques presented here would provide not only a Hilbert space description and the vacuum states of the models, but also the operator algebra generating and measuring the excitations.

Recently a new class of 4D TQFT's has been constructed by B\"arenz and Barrett \cite{BB}. In this construction a handle decomposition of the 4D manifolds in the form of Kirby calculus \cite{kirby} is instrumental. In fact with the Heegaard splitting we also used a variant of a handle decomposition of a 3D manifold. We believe that the ideas presented here could allow to construct a Hilbert space representation of the 4D TQFT's presented in \cite{BB} and furthermore generalize these TQFT's to TQFT's with defects. 

Barrett \cite{privatecomm} also proposes that interacting theories can be constructed from TQFT's by considering very complicated topologies. Here we see an interesting version of this proposal realized. Degrees of freedom are added to a TQFT by allowing defects. This defect structure is however encoded into an intricate structure of the topology of a lower dimensional manifold, here the Heegaard surface.  Note that there are two dimensional reductions: first from the 4D space time manifold to the 3D `spatial' hypersurface, then from the 3D manifold to the 2D Heegaard surface. A further topic of future research will be the construction of time evolution: This can either involve a fixed 3D triangulation evolving in time, or, as discussed in \cite{DittHoe1,DittHoe2,DittStein,H1,H2} an evolution with a time varying triangulation.  The latter would mean that the topology of the Heegaard surface changes during (discrete) time evolution. Different possibilities also exist for the type of dynamics: one can either implement a 4D TQFT  with non--interacting defects.  In this case the dynamics would be given by further constraints (or stabilizers): these would require flatness for all edges of the triangulation not carrying a defect and prescribe the curvature around the edges carrying a defect. See for instance \cite{BaezAlex,Fairbairn,thooft} for the exploration of BF theory with string--like defects.  Alternatively one can aim at an interacting theory, such as 4D gravity. In fact we hope that the BF representation constructed in \cite{DG14a,DG14,BDG15} as well as the Turaev-Viro representation \cite{DG16} and its possible generalization to $(3+1)$D based on the work presented here, will give an interesting starting point for constructing the dynamics of 4D quantum gravity.

\begin{appendix}

\section{ Refining of the graph, adjusted to a ribbon} \label{refiningG}

\begin{figure}[h!]
	\centering
	\begin{minipage}[b]{0.40\textwidth}
		\centering
		\includegraphics[scale = 1]{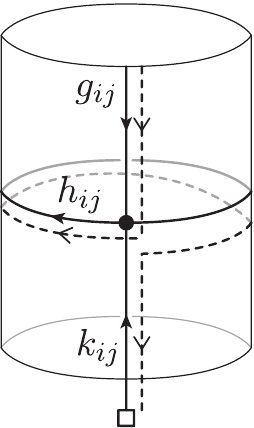}	
	\end{minipage}
	\begin{minipage}[b]{0.40\textwidth}
		\centering
		\includegraphics[scale = 1]{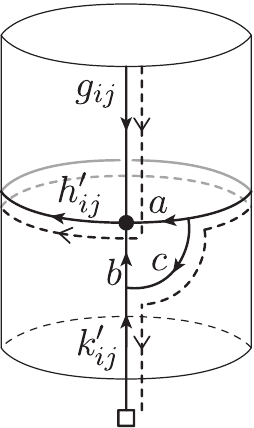}	
	\end{minipage}
	\caption{Example of a ribbon operator that requires a refining of the graph. The bold-solid line represents the graph on the boundary of the tube while the dashed line represents the path $\alpha$ followed by the ribbon. The small square represents the target node of $\alpha'$. Each time the ribbon crosses a link $l$, it acts on it. In the situation depicted on the left, the ribbon crosses the links labeled by $h_{ij}$ and $k_{ij}$ which are also part of the shadow path $\alpha'$. To circumvent this difficulty, we refine the graph and introduce auxiliary holonomies so that the path $\alpha'$ associated with the holonomy $(k_{ij}')^{-1}c\,h_{ij}'g_{ij}$ avoids the links crossed by $\alpha$. \label{refG}}
\end{figure}

In some cases  we need also ribbon operators that cross links $l$ that are however also part of the shadow path $\alpha'$. Such a situation in depicted in figure \ref{refG}.    This situation can be resolved by introducing auxiliary links with associated auxiliary holonomy variables. In introducing this auxiliary variables we use the fact that we deal with a locally flat connection, and consider gauge invariant functions of the holonomies. Thus the auxiliary variables will not add any further information. In fact, in the end we will be able to express the action of the ribbon without making use of the auxiliary variables.

The situation depicted in figure \ref{refG} will arise for the example we consider in section \ref{4simplex}. We subdivide the two links that are supposed to be crossed by the ribbon and connect the two new nodes by a new auxiliary link. 
The new holonomies are related to the original ones as follows (see figure \ref{refG} for the notation):
\ba
h_{ij}\,=\, a h'_{ij}  \q , \q k_{ij} \,=\, b k'_{ij} \q , \q \mathbb{I}\,=\, a^{-1} b c \q .
\ea

 We can now define a ribbon which crosses the subdivided links but where the path $\alpha'$ now avoids all links crossed by the ribbon, or more precisely $\alpha$. 
We can work out the action of the ribbon (parametrized by a translation $H$ and w.l.o.g. ending at the source node of the link belonging to $k_{ij}$) on these holonomies: 
\ba
a \q  \rightarrow \q &&[(h'_{ij})^{-1} c^{-1} k'_{ij}] H^{-1} [(k'_{ij})^{-1} c h'_{ij}]\, a \nn\\
b \q \rightarrow \q && [(h'_{ij})^{-1} c^{-1} k'_{ij}] H^{-1} [(k'_{ij})^{-1} c h'_{ij}]\, b \q .
 \ea
Thus the original variables are shifted by
\ba
h_{ij} \q  \rightarrow \q &&[(h_{ij})^{-1} k_{ij}] H^{-1} [(k_{ij})^{-1}  h_{ij}]\, h_{ij} \nn\\
k_{ij} \q \rightarrow \q && [(h_{ij})^{-1}  k_{ij}] H^{-1} [(k_{ij})^{-1}  h_{ij}]\, k_{ij} \q .
\ea
We have indeed in the expression for the parallel transport (in square bracket) for the shift parameter $H$, the shifted variables $h_{ij}$ and $k_{ij}$ itself appearing. But this happens in a combination, that is invariant under the shift. Thus the action of the ribbon is well defined.

\end{appendix}

\begin{center}
\textbf{Acknowledgements}
\end{center}
BD would like to thank John Barrett, Marc Geiller, Wojciech Kaminski and Aldo Riello for very useful discussions. We thank Aldo Riello for a careful reading of a draft of the paper. CD's work is supported by an NSERC Discovery grant awarded to BD. This work is supported by Perimeter Institute for Theoretical Physics. Research at Perimeter Institute is supported by the Government of Canada through Industry Canada and by the Province of Ontario through the Ministry of Research and Innovation.

\end{document}